\documentclass[twocolumn,secnumarabic,amssymb,nobibnotes,aps,pre]{revtex4-1}
\usepackage{graphicx} 
\usepackage{amsmath}
\usepackage{xcolor}
\usepackage{graphicx}
\usepackage{dcolumn}
\usepackage{bm}
\usepackage{amssymb}
\usepackage{float}
\usepackage{graphicx}		 
\usepackage{amsmath,eqnarray}
\newcommand{\fl}{\hspace*{-0.5cm}}
\usepackage{subfig}
\usepackage{float}
\usepackage{multirow}

\begin{document}

\title{Mechanisms for extreme events in position-dependent mass systems}

\author{Wasif Ahamed M$^{1,2}$, Chithiika Ruby V$^{3,4}$, Sathish Aravindh M$^{2}$, Venkatesan A$^{1}$, Lakshmanan M$^{2}$}

\affiliation{\footnotesize $^{1}$  PG \& Research Department of Physics, Nehru Memorial College (Autonomous), Affiliated to Bharathidasan University, Puthanampatti, Tiruchirappalli 621 007, Tamil Nadu, India.}

\affiliation{\footnotesize $^{2}$Department of Nonlinear Dynamics, School of Physics, Bharathidasan University, Tiruchirappalli - 620 024, Tamil Nadu, India.}

\affiliation{\footnotesize $^{3}$Center for Nonlinear and Complex Networks, SRM TRP Engineering College, Tiruchirappalli - 621 105, Tamil Nadu, India.}

\affiliation{\footnotesize $^{4}$Center for Research, Easwari Engineering College, Chennai - 600 089, Tamil Nadu, India.}

\email{ahamedw019@gmail.com, \\
chithiikaruby.v@trp.srmtrichy.edu.in, \\
sathisharavindhm@gmail.com, \\ av.phys@gmail.com,  \\ lakshman.cnld@gmail.com}

\begin{abstract}
Oscillators defined on curved spaces provide a natural framework for exploring geometry-induced nonlinear dynamics. The two-dimensional Mathews–Lakshmanan oscillator and Higgs oscillator are prominent examples of geometrically induced nonlinear oscillators. The two-dimensional Higgs oscillator in flat space arises from a conformal (stereographic) projection of the harmonic oscillator on the sphere $\mathcal{S}^2$ onto the two-dimensional plane. Its one-dimensional counterpart, defined by a non-Euclidean geometry with curvature parameter, exhibits chaotic dynamics and extreme events (EEs) when subjected to damping and external driving forces \cite{wasif2025extreme}. In this work, we introduce coupling through mass interaction among multiple one-dimensional damped and forced Higgs oscillators and investigate the resulting collective nonlinear dynamics within a many-particle framework. More importantly, we establish the connection between the conformal mass $m(x)$ and the extrema of velocity through a power-law relation, demonstrating how the conformal mass regulates the turning points leading to the emergence of extreme events (EEs). Furthermore, to provide additional insight into the underlying mechanism, we quantify the energy synchronization error among the oscillators, thereby revealing how energy synchronization evolves during the onset of EEs.
\end{abstract}

\maketitle

\section{Introduction}

Extreme events (EEs), like rogue waves, earthquakes, and stock market crashes, spontaneously emerge in complex nonlinear dynamical systems as emergent properties \cite{lucarini2016extremes, albeverio2006extreme, chowdhury2022extreme}. Understanding and predicting these events and their unique statistics remain a major challenge across various areas of applied sciences, requiring interdisciplinary research. A dynamical perspective suggests that even normal fluctuations can generate these extreme excursions, with the field incorporating time-series analysis, generalized extreme value theory and deterministic modeling \cite{chowdhury2022extreme}. Identifying the underlying mechanisms of EEs, such as crises in chaotic attractors, is crucial for ensuring dynamical stability and improving predictability. During the onset of EEs, the attractor undergoes an intermittent expansion caused by the dynamical crises, such as an interior crisis, where the attractor collides with an unstable invariant set within its basin of attraction, or a boundary crisis at the basin boundary \cite{grebogi1983crises, grebogi1987critical}. The resulting intermittent long excursions are classified as EEs only when they exceed a predefined statistical threshold, which occur as rare but recurrent events through intermittent visits to crises regions in the phase space and producing a statistically significant tail in the distributions of local maxima or threshold exceedance, consistent with the principles of extreme value theory \cite{lucarini2016extremes, albeverio2006extreme}.

Similarly, several other mechanisms, such as Pomeau - Manneville intermittency \cite{pomeau2017intermittent} and attractor-merging crises \cite{grebogi1983crises, grebogi1987critical}, have also been explored. The occurrence of extreme events has been reported in a wide range of dynamical systems, including Li\'{e}nard systems \cite{kingston2017extreme}, FitzHugh - Nagumo oscillators \cite{saha2017extreme}, Hindmarsh--Rose models \cite{saha2018riddled, vijay2023superextreme}, electronic circuits \cite{kingston2017extreme, thangavel2021extreme}, networks of Josephson junctions \cite{ray2020extreme}, the nonlinear Schr\"{o}dinger equation \cite{fotopoulos2020extreme} and dynamical systems with discontinuous boundaries \cite{kumarasamy2018extreme}.

Within the class of mechanical oscillators, investigations of extreme events have predominantly focused on systems with finite-order polynomial nonlinearities, such as the Rayleigh \cite{kaviya2022extreme}, Helmholtz \cite{sudharsan2025extreme} and parametrically driven Duffing oscillators \cite{zhao2022extreme} with the influence of linear drag and external forcing. More recently, considerable attention has shifted toward mechanically relevant systems exhibiting non-polynomial nonlinearities, including particles confined in rotating parabolic potentials\cite{sudharsan2021emergence}, diatomic Morse oscillators \cite{durairaj2023emergence}, nanomechanical oscillators and most recently, nano-electromechanical resonators \cite{venkatesh2026superextreme}, whose dynamics can be viewed as a generalization of the Mathews - Lakshmmanan nonlinear oscillator. Motivated by these developments, in this manuscript we investigate the emergence of extreme events in the nonlinear Higgs oscillator, a non-polynomial oscillator that naturally arises from field-theoretic considerations and the dynamics of particles on spaces with constant curvature.

The Higgs oscillator was introduced as the dynamics of a harmonic oscillator defined on a space of constant curvature \cite{higgs1979dynamical}. In another way, it can also be viewed as the zero-mode reduction of a  $SU(2) X SU(2)$-invariant Lagrangian density representing the ultraviolet behaviour of meson fields \cite{delbourgo1969infinities}. The interaction Lagrangian density for that meson field, $\vec{\phi}$, in Schwinger coordinates takes the nonlinear form as 
\begin{equation}
\fl {\cal L}_{int} = - \frac{\kappa}{2\;(1 + \kappa \phi^2)}\left(\phi^2(\partial_{\mu} \vec{\phi}).(\partial_{\mu} \vec{\phi}) + \frac{(\vec{\phi}.\partial_{\mu} \vec{\phi})(\partial_{\mu} \vec{\phi}.\vec{\phi})}{1 + \kappa\;\phi^2}\right). \label{Lint}
\end{equation}

We consider an extended Lagrangian density obtained by adding a quadratic interaction term,  $\phi^2$, to the original Lagrangian density, which can be written as  
\begin{eqnarray}
\fl {\cal L} & =& \frac{1}{2} (\partial_{\mu } \vec{\phi}).(\partial_{\mu} \vec{\phi})\\ 
&-& \frac{\kappa}{2\;(1 + \kappa \phi^2)}\left(\phi^2(\partial_{\mu} \vec{\phi}).(\partial_{\mu} \vec{\phi}) + \frac{(\vec{\phi}.\partial_{\mu} \vec{\phi})(\partial_{\mu} \vec{\phi}.\vec{\phi})}{1 + \kappa \phi^2}\right) - \phi^2. \nonumber \\
\end{eqnarray}

Under the zero-mode reduction, $\vec{\phi}(x,t)\rightarrow\vec{q}(t)$, the infinite-dimensional field theory reduces to a finite-dimensional Higgs oscillator as
\begin{eqnarray}
\fl \quad L &=& \frac{1}{2}\left[\dot{\vec q}^{2}
-\frac{\kappa}{(1+\kappa q^2)}
\left(q^2\dot{\vec q}^{\,2}
+\frac{(\vec q\cdot\dot{\vec q})^2}{1+\kappa q^2}\right)\right]
-q^2. \label{Lagrangian_Higgs}
\end{eqnarray}
The connection between the nonlinear oscillator and nonlinear field theories broadens the scope of applications of the Higgs oscillator. In particular, it establishes its relevance in condensed matter physics, such as in the continuum description of Heisenberg spin chains \cite{daniel1992singularity} and in high-energy physics, including effective field-theoretic studies of quark dynamics \cite{Gasser1984142}.  

Alternatively, the Lagrangian of the Higgs oscillator (\ref{Lagrangian_Higgs}) is non-polynomial in the dynamical variables and can be equivalently interpreted as a position-dependent mass type system \cite{ruby2024lienard}. In this formulation, the effective mass is not introduced phenomenologically but arises naturally from the geometry of the underlying curved configuration space through the curvature parameter $\kappa$.  Exploring nonlinear phenomena in such position-dependent mass systems may broaden their applicability to semiconductor heterostructures \cite{bastard1989wave, gora1969theory},  optical media \cite{Khordad2011} and quantum dots \cite{ Alaakol2026}. Rare velocity bursts in these systems may provide insights into anomalous transport phenomena, enhanced energy localization, or unusually rapid propagation of collective excitations. The present study therefore offers a minimal geometric mechanism for understanding how curvature-induced nonlinearities can generate extreme events under external driving or dissipative conditions.

The mechanisms leading to extreme events (EEs) in isolated nonlinear dynamical systems have been largely investigated \cite{chowdhury2022extreme}. In contrast, the processes responsible for the emergence of EEs in coupled systems remain relatively underexplored. Consequently, understanding how extreme events arise in coupled nonlinear dynamical systems has become a central focus of interdisciplinary research in recent years \cite{ansmann2013extreme, ray2020extreme, roy2024impact,ardhanareeswaran2025intermittent}. Significant progress has been reported in this area including studies on EEs in networks of FitzHugh–Nagumo oscillators as well as in globally coupled Josephson junctions and Liénard-type oscillators \cite{ansmann2013extreme,ray2020extreme}. Sudharsan and co-workers investigated the emergence of extreme events in a system of two diffusively and bidirectionally coupled R\"{o}ssler oscillators \cite{sudharsan2025extreme}. More recently, Shashangan et al. examined the propagation of extreme events across two distinct layers of a multiplex network \cite{shashangan2025propagation}. Additionally, time-varying interactions and dynamic rewiring of network connections have been identified as precursors to such events. Other structural changes in the network, such as the inclusion of time-delayed couplings or the interaction of counter-rotating oscillator pairs, have also been shown to trigger the onset of extreme events \cite{saha2017extreme, varshney2021traveling}.

In this manuscript, we believe that to the best of our knowledge it is for the first time we model the coupled Higgs oscillators with mass-like interaction and unveil the mechanism for the emergence of EEs by establishing certain power law relation  between the maxima of the velocity profile time series, say $y_{max}$ to the conformally-induced mass function, say $m(x)$, thereby linking the extreme velocity, say $y_{EE}$, to the mass, $y_{EE} \subset y_{max}$. Previously, several authors have studied extreme events in polynomial oscillators \cite{kingston2017extreme, saha2017extreme, suresh2018influence, chowdhury2022extreme}. However, the exploration of extreme events in non-polynomial oscillators is particularly important, as many physical systems exhibit oscillatory behavior that cannot be described by polynomial functions. Examples include pendulums, mechanical systems, electronic circuits and certain biological rhythms \cite{chowdhury2022extreme}. In a recent study, we observed such complex dynamics in the one-dimensional Higgs oscillator, a non-polynomial oscillator, when subjected to damping and external driving forces \cite{wasif2025extreme}. Motivated by these results, we aim to extend our investigation to many-particle dynamics, starting with two independent Higgs oscillators that interact locally with each other and then extend the study to a large number of coupled oscillators. We hope such studies are valuable for predicting dynamical behavior like the position dependent effective mass of the electron $m^{*}=\hbar^{2}/{\frac{d^{2}E}{dk^{2}}}$ in the heterogeneous band structures \cite{PhysRevB.16.1675}. Also, this study will open the arena for exploring the non-integrable classical analogue of an interacting position-dependent mass system and helps one to check the intactness of the quantum dynamics of interacting position dependent masses. Since this study unveils the relations between extrema velocity with position dependent conformal mass, it can be utilized to optimize system performance. The key insight is that an appropriate transformation of the phase-space variables enables a distinct visualization of the instability region in the transformed space, thereby facilitating the prediction of extreme events (EEs).

The paper is organized as follows. In Section II, we present the mathematical model and derive the governing equations of motion of the coupled Higgs oscillator system. Section III investigates the dynamical behavior of the system through bifurcation diagrams and the corresponding Lyapunov exponents. Subsequently, the temporal evolution is analyzed and a power-law relationship between bounded chaos and extreme events is established. The influence of the forcing amplitude and the curvature constant is then examined through a two-parameter phase diagram, which delineates the regions of extreme and non-extreme events, together with their corresponding energy synchronization error. The proposed power-law scaling is further validated using an early warning indicator for the prediction of extreme events. Section IV extends the analysis to a network of $N$-coupled Higgs oscillators. In this section, the degree of extreme events is quantified over a broad range of interaction strengths and supported by the average energy synchronization error. Furthermore, the combined effects of the forcing amplitude and the curvature constant are investigated, followed by a detailed analysis of the temporal dynamics, statistical characteristics and validation of the proposed power-law relationship. Finally, Section V summarizes the main findings and presents the concluding remarks.

\section{Model}

\begin{figure*}[]
    \centering
    \includegraphics[width=0.95\linewidth]{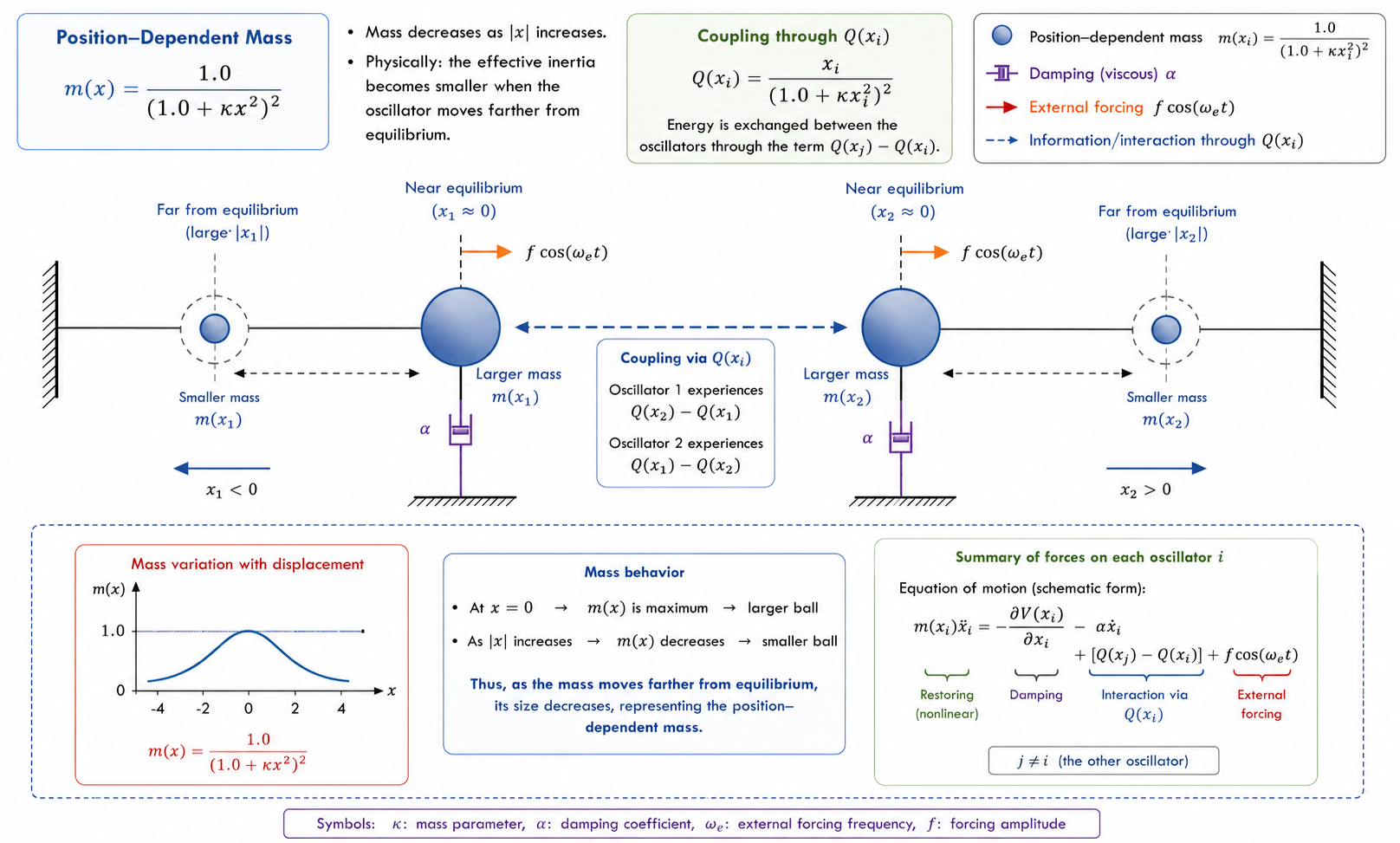}
\caption{The schematic diagram shows the  position dependent Higgs oscillators interacting with each other through nonlinear mass interaction ($Q(x_{j}) -Q(x_{i})$) under the influence of linear damping ($\alpha y $) and external forcing $(fcos(\omega_{e}t))$. Note : The schematic diagram was originally hand-drawn and subsequently processed using an AI image-generation tool for coloring and layout enhancement.} 
    \label{SD}
\end{figure*}

The nonlinear oscillators introduced by Mathews - Lakshmanan \cite{mathews1974unique, mathews1975quantum} and by Higgs \cite{higgs1979dynamical} exhibit an amplitude-dependent simple harmonic oscillatory behavior but take intrinsically distinct forms when expressed in flat-space coordinates. When the geodesic coordinates of the harmonic oscillator on a sphere are mapped to the Euclidean plane through gnomonic projection, which preserves the full spherical symmetry, the resulting flat-space system is in a nonlinear, non-polynomial form known as the Higgs oscillator \cite{carinena2004non}. The Lagrangian of the one-dimensional version takes the form, 
\begin{equation}
L = \frac{\dot{x}^{2}}{2\,(1+\kappa\;x^{2})^{2}} - \frac{\omega_{0}^{2}x^{2}}{2}, \label{1D-HO}
\end{equation}
where $\kappa$ encodes the space curvature and $\omega_{o}$ is the natural frequency of the Higgs oscillator.

In contrast, if a harmonic oscillator on a sphere embedded in a higher-dimensional flat space is mapped to a Euclidean subspace by orthogonal projection, the resulting system corresponds to the Mathews–Lakshmanan oscillator. The corresponding one-dimensional Lagrangian takes the form 
\begin{equation}
L = \frac{\dot{x}^{2} - \omega_{0}^{2}x^{2}}{2\,(1+\kappa x^{2})},
\end{equation}
where the parameters, $\kappa$ and $\omega_{0}$, are the radius of the curvature and the natural frequency of the system ~\cite{mathews1974unique,mathews1975quantum}.  

Both the above systems are examples of position-dependent mass models but possess different mass functions as they result from different types of projection \cite{ruby2024lienard, ruby2021classical}. These systems have subsequently been generalized and analyzed from various perspectives. For example, the classical and quantum studies on the Mathews-Lakshmanan oscillator had been extended to $3$-dimensional \cite{lakshmanan1975quantum} and $d$-dimensional generalizations  \cite{ranada2002harmonic, lakshmanan2013generating} and to rational extensions of the potentials  \cite{quesne2018deformed}. Analogously, the Higgs oscillator has been studied in different contexts including superintegrability \cite{ballesteros2008superintegrable, ruby2021classical}, relativistic generalizations \cite{mohammadi2016dirac}, exact quantum solvability \cite{carinena2004non}, hidden symmetries and the associated conformal algebra \cite{evnin2016ads} and construction of nonlinear coherent states related to deformed oscillator algebras \cite{mahdifar2006geometric, ruby2025algebraic}. 

\begin{figure}[]
    \centering
     \includegraphics[width=0.77\linewidth]{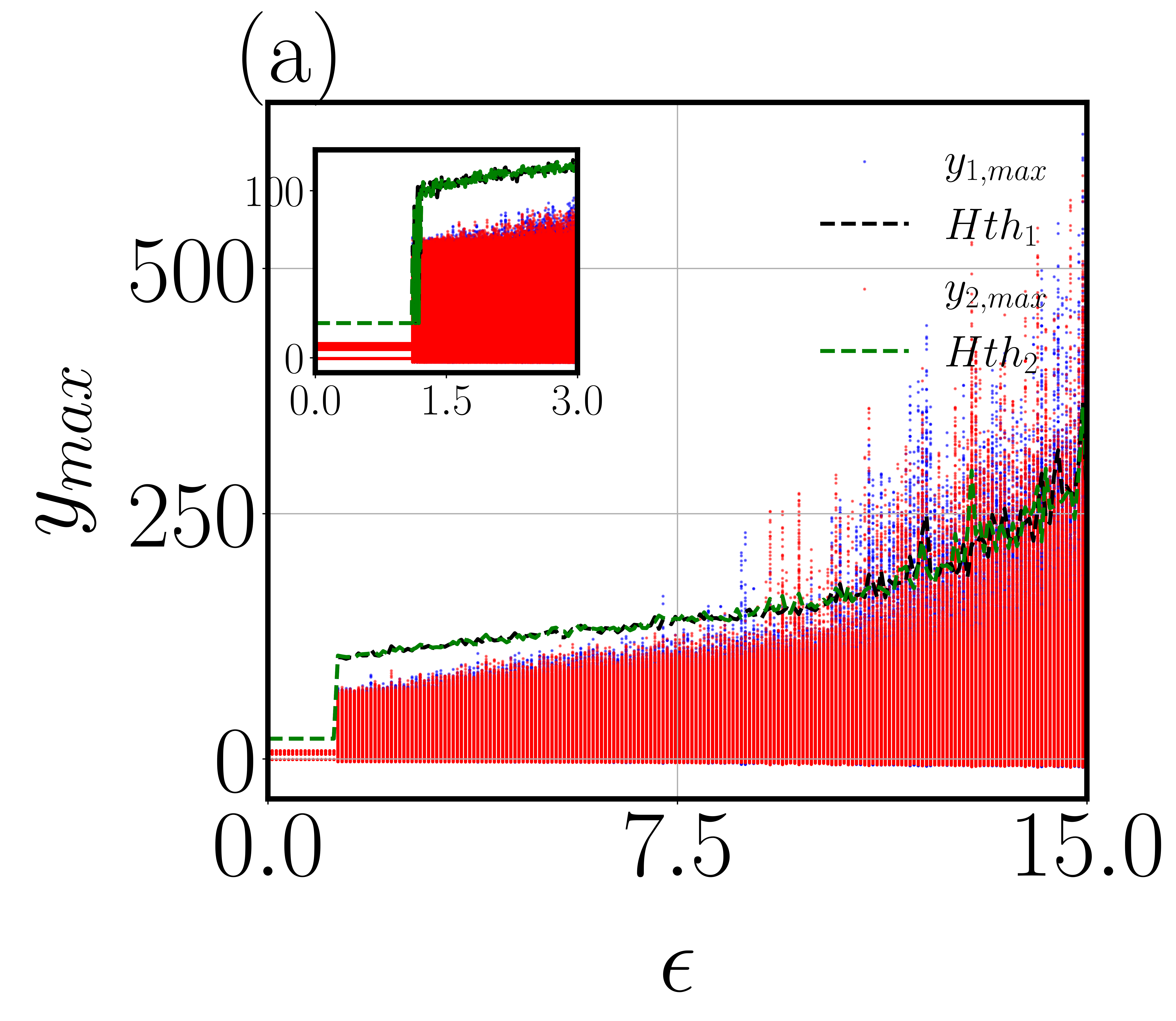}
   \includegraphics[width=0.77\linewidth]{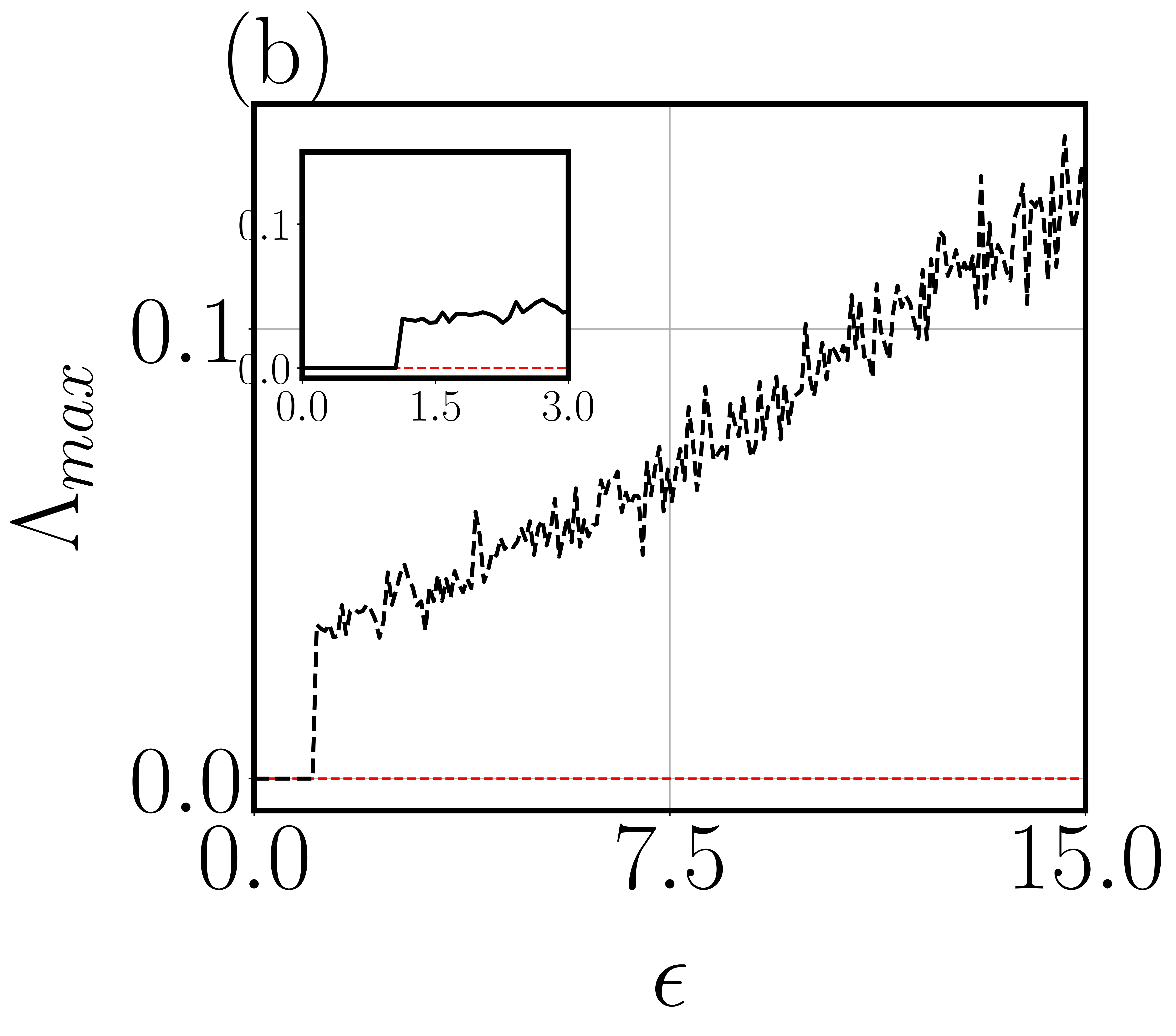}
\caption{Panel (a) shows the bifurcation plot illustrating the transition from a quasi-periodic regime to a chaotic regime in which extreme events emerge as the coupling strength between the oscillators is varied over the range $\epsilon \in (0.0,15.0)$, with fixed values of $\kappa=0.21$, $\omega_{o}^{2}=0.1$, $f=4.3251$, and $\omega_{e}=0.1495$. The blue and red dots in the bifurcation diagram represent the peak velocities, $y_{1,\max}$ and $y_{2,\max}$, of oscillators 1 and 2, respectively. Panel (b) shows the corresponding maximal Lyapunov exponent characterizes the underlying dynamics associated with the bifurcation structure.}
    \label{bif-lya}
\end{figure}

A rich variety of nonlinear phenomena such as period-doubling bifurcation preceded by symmetry-breaking bifurcations, chaos, intermittency, anti-monotonicity and torus doubling followed by a strange non-chaotic attractor had been observed in a non-polynomial nonlinear system \cite{venkatesan1997nonlinear}.  In this context, the one-dimensional Higgs oscillator (\ref{1D-HO}) has also been studied and its equation of motion is given by 
\begin{equation}	
\ddot{x}-\dfrac{2\kappa x}{(1+\kappa x^2)}\dot{x}^2+\omega_0^2(1+\kappa x^2)^2 x=0. 
\label{equ1}
\end{equation}
It can be exactly integrated and  its solution shows simple oscillatory behavior
\begin{equation}
x(t)=\dfrac{A \sin(\Omega t + C)}{\sqrt{1- {\kappa} A^2 \sin^2(\Omega t + C)}},
\label{equ5}
\end{equation}
where $A$ and ${\displaystyle \Omega\;\bigg(=\frac{\omega_{o}}{\sqrt{1-\kappa A^{2}}}\bigg)}$ are the amplitude and frequency of the oscillator \cite{ruby2021classical}. Hence, $x(t)$ is periodic for all the values of $ A $ for $ \kappa < 0 $ and periodic in the range $ |A|<\dfrac{1}{\sqrt{\kappa}} $ when $ \kappa > 0 $. 

\begin{figure*}[]
    \centering
    \includegraphics[width=0.30\linewidth]{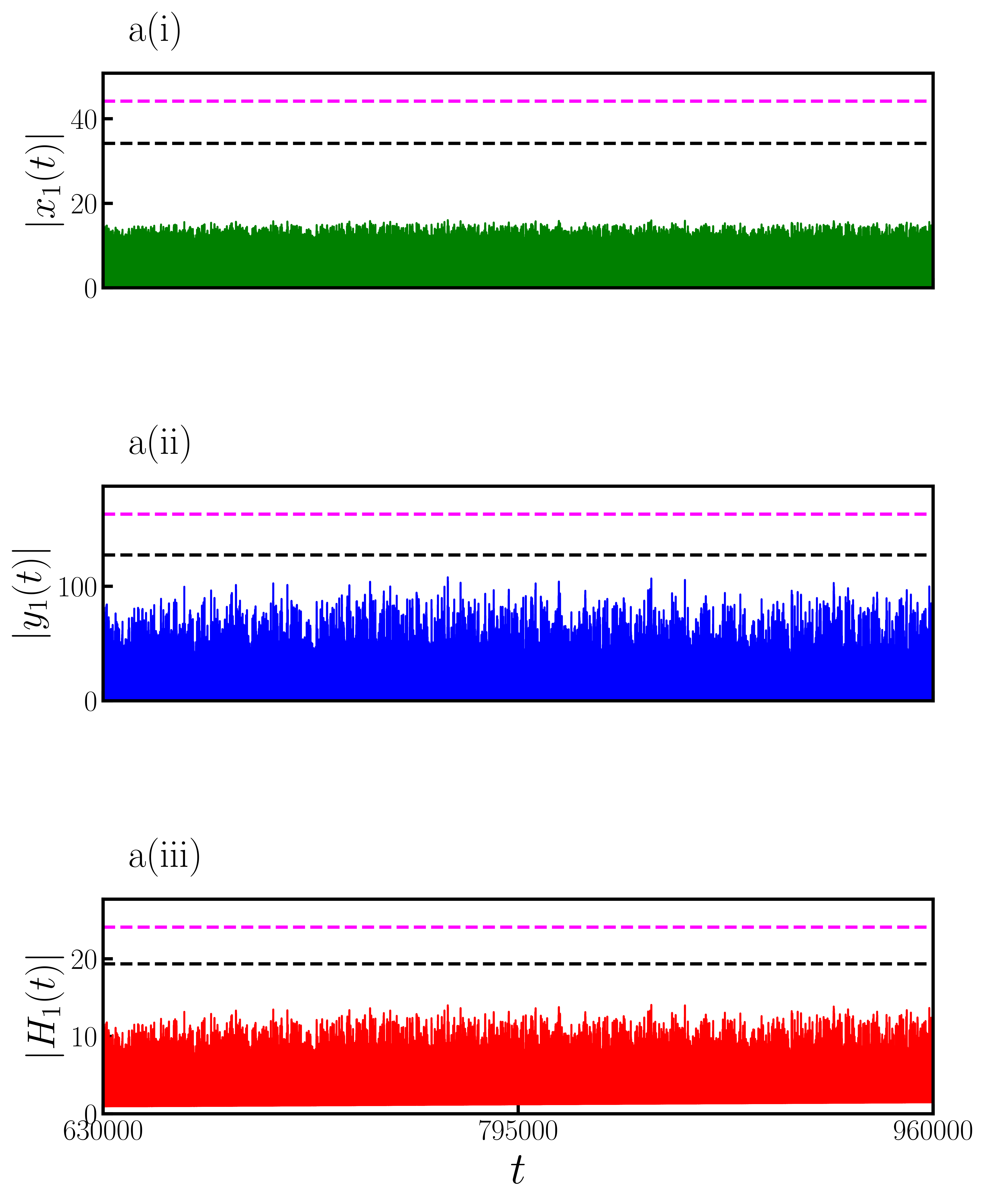}
    \includegraphics[width=0.30\linewidth]{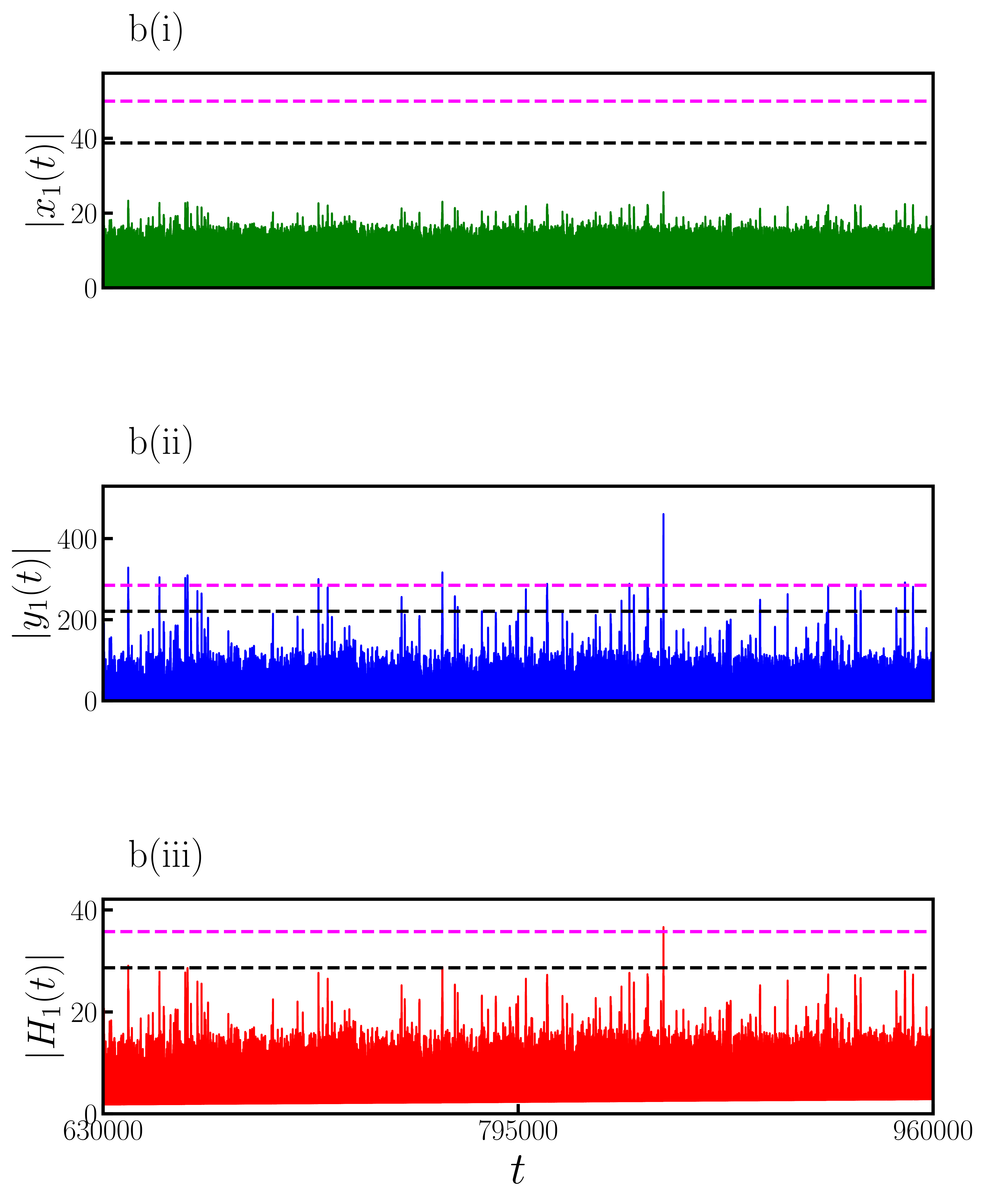}
    \includegraphics[width=0.34\linewidth]{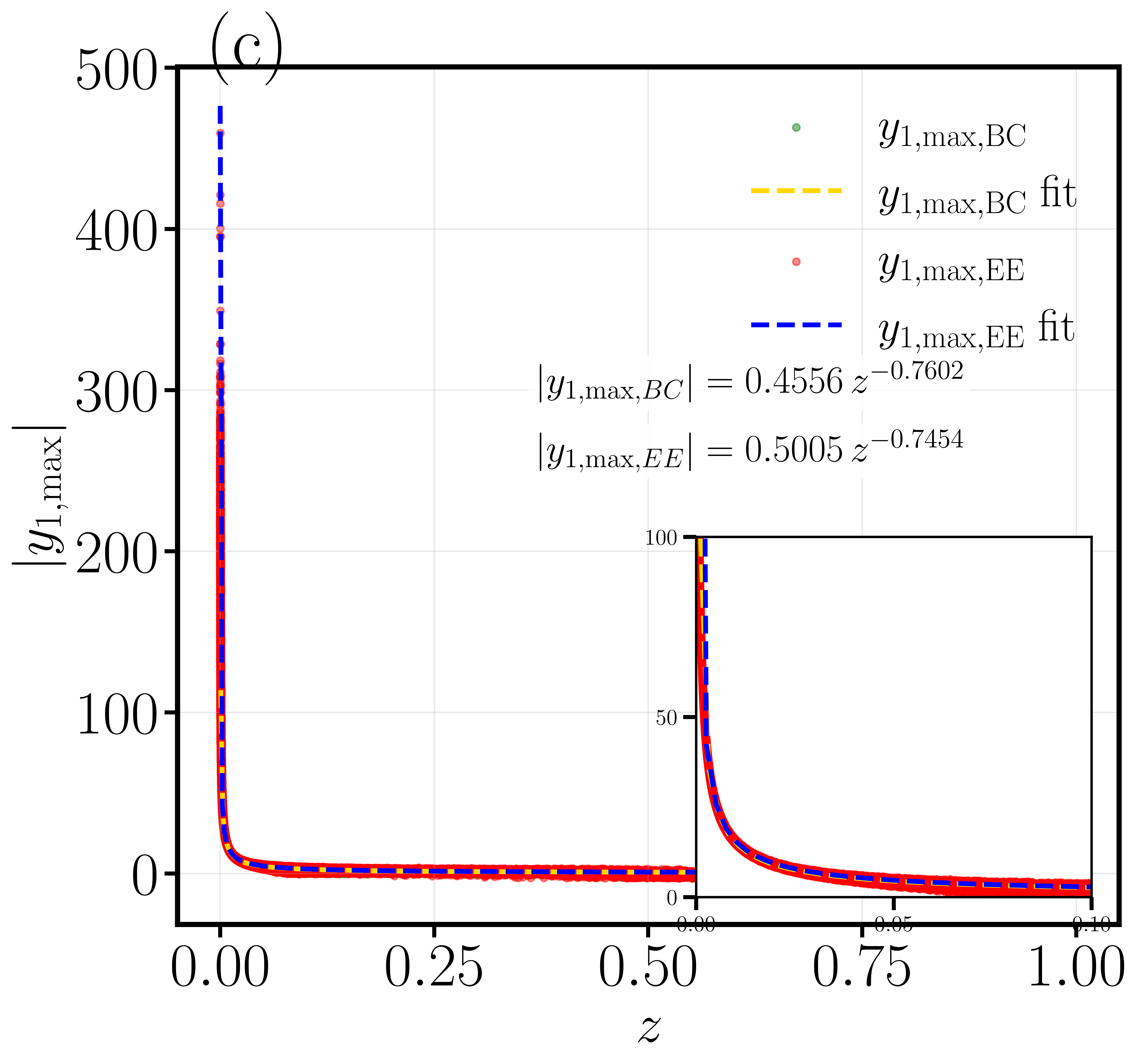}
    \includegraphics[width=0.30\linewidth]{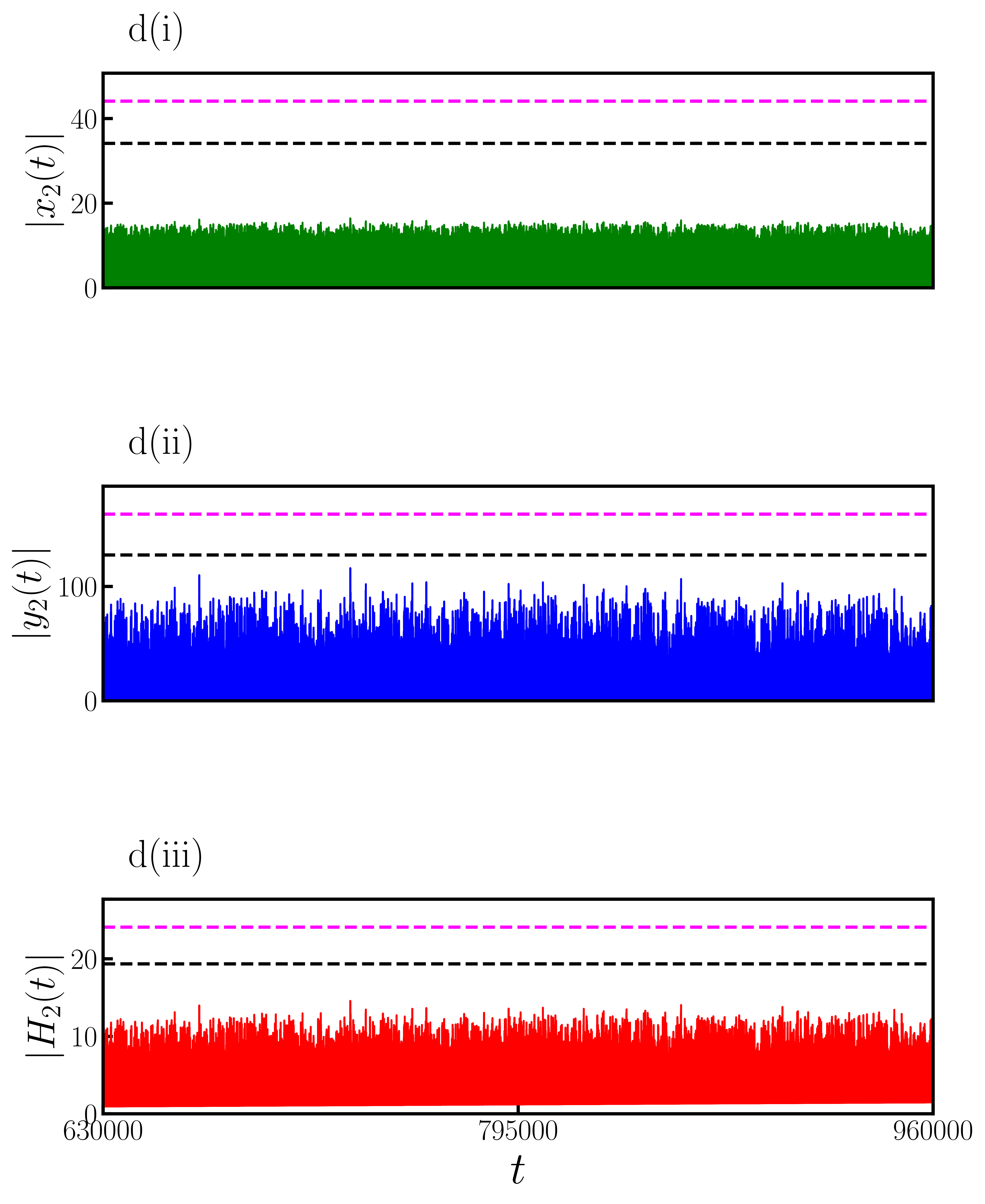}
     \includegraphics[width=0.30\linewidth]{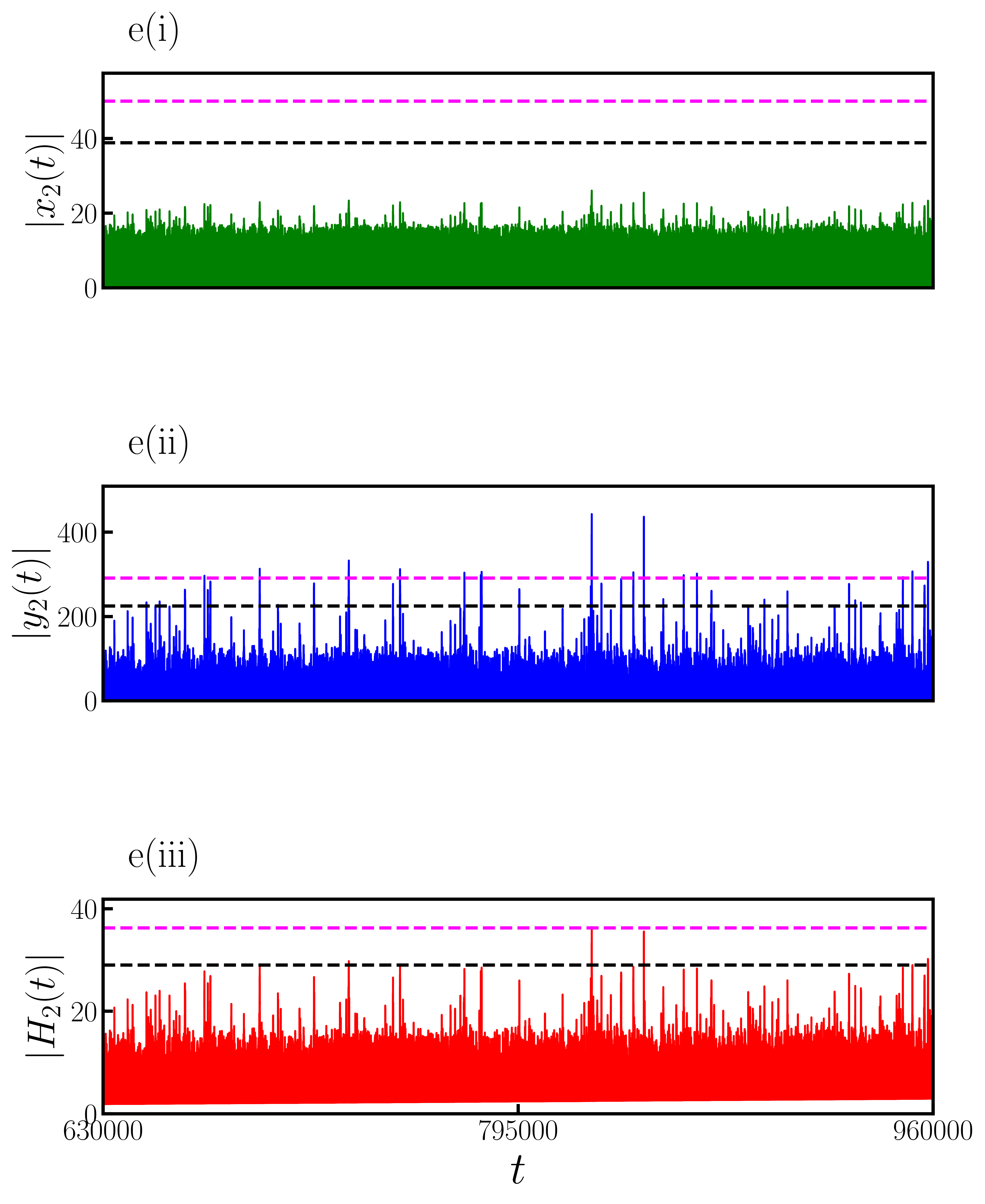}
     \includegraphics[width=0.34\linewidth]{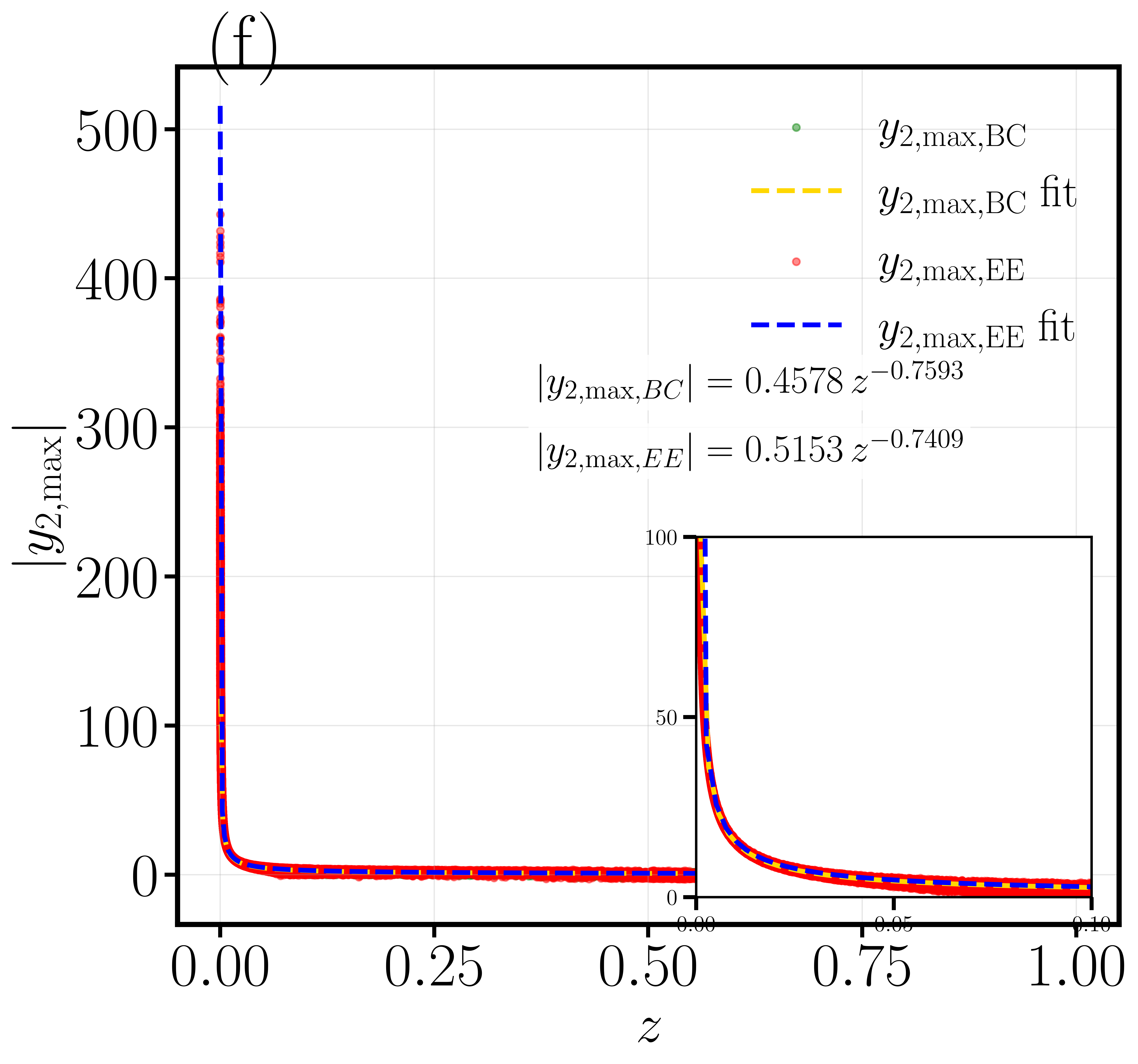}
   \caption{Panel (a(i-iii)) and (d(i-iii)) show the bounded chaotic (BC)  time series profile of position(green), velocity(blue) and the instantaneous energy (red) of oscillators 1 and 2, respectively. Panel (b(i-iii)) and (e(i-iii)) display the chaotic time series  profile of position(green), velocity(blue) and the instantaneous energy,  where the velocity exhibits extreme events (EEs), for oscillators 1 and 2, respectively. Panel (c) and (f) depict the relation between the conformal mass factor $(m(x))$  and the absolute maximum velocity, i.e., $( |y^{1,2}_{\text{max}}|)$, along with their power-law fits.
}
    \label{TSPWL}
\end{figure*}
However, it is reported that the system (\ref{equ1}) departs from its natural oscillatory motion (\ref{equ5}) when subjected to damping and driving forces. Its dynamics is governed by the equation 
\begin{equation}
\ddot{x}-\dfrac{2\kappa x}{(1+\kappa x^2)}\dot{x}^2+\omega_0^2(1+\kappa x^2)^2 x+\alpha \dot{x}  = f \cos \omega t.
\label{equ6}
\end{equation}
It exhibits typical nonlinear phenomena such as bifurcations, chaos and extreme events. Here $\alpha$ is the damping parameter, $f$  and $\omega$ are the amplitude and frequency of the external driving force \cite{wasif2025extreme}. 
\begin{figure}[h!]
    \centering
    \includegraphics[width=0.45\linewidth]{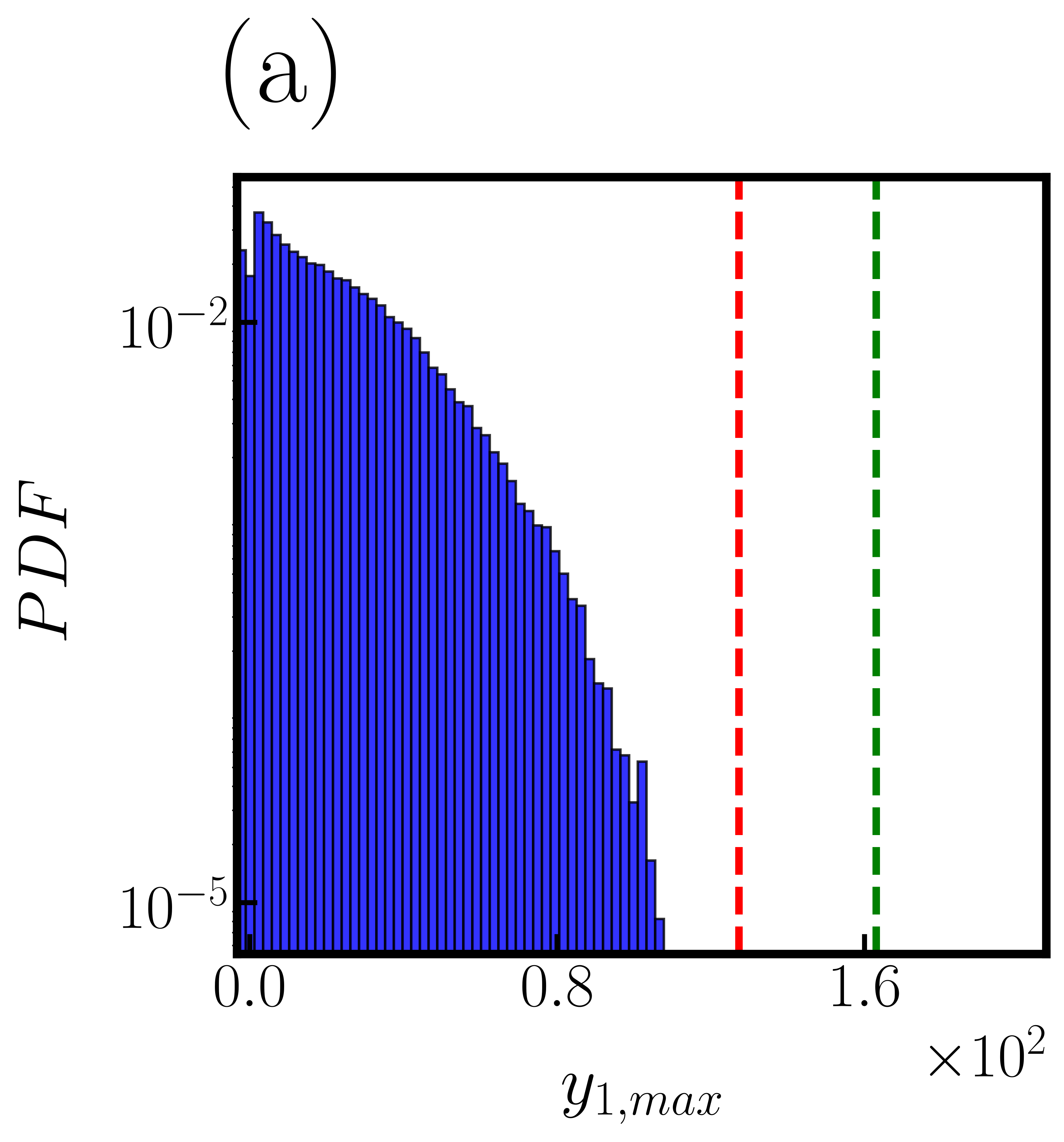}
    \includegraphics[width=0.45\linewidth]{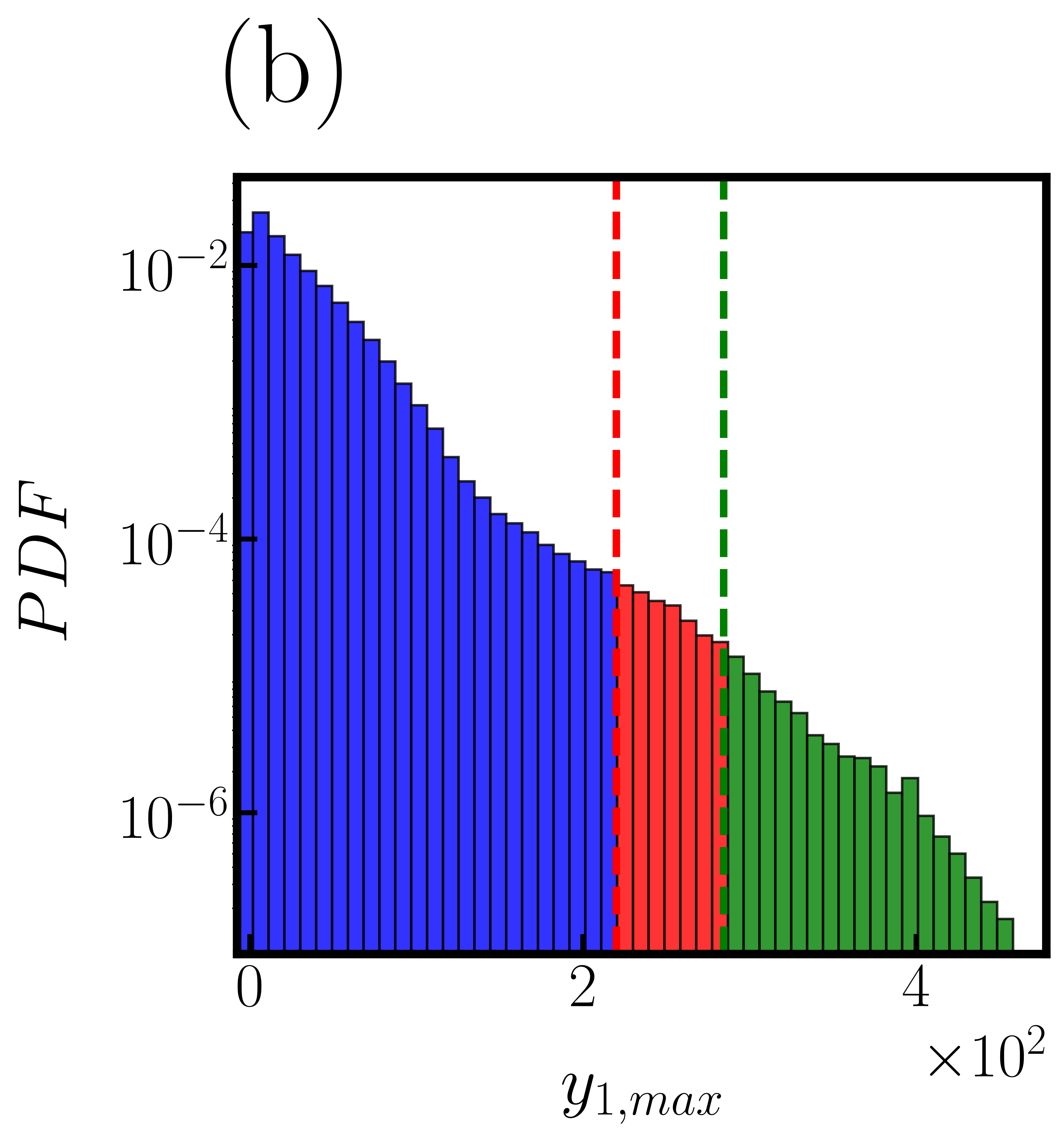}
    \includegraphics[width=0.45\linewidth]{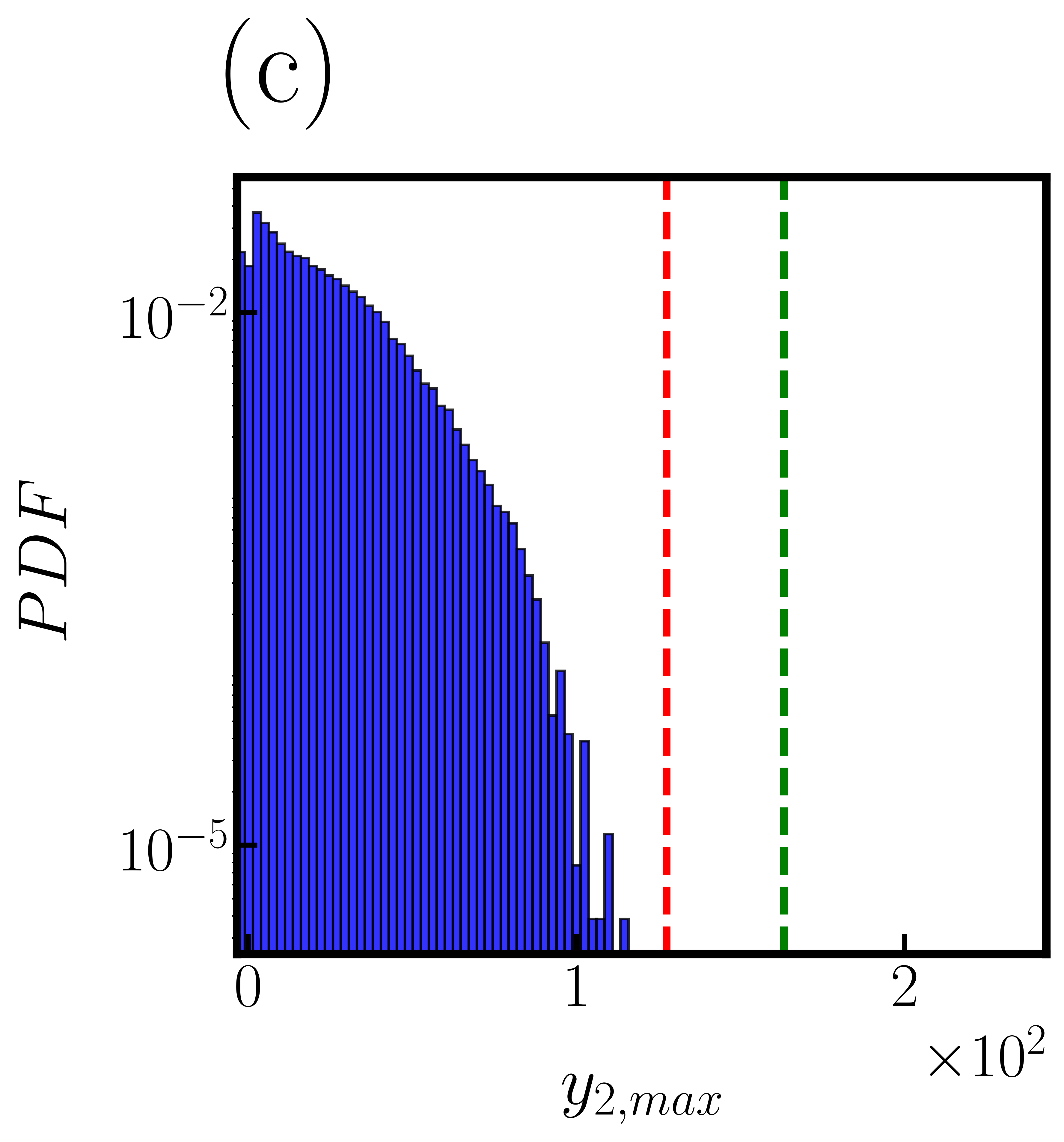}
    \includegraphics[width=0.45\linewidth]{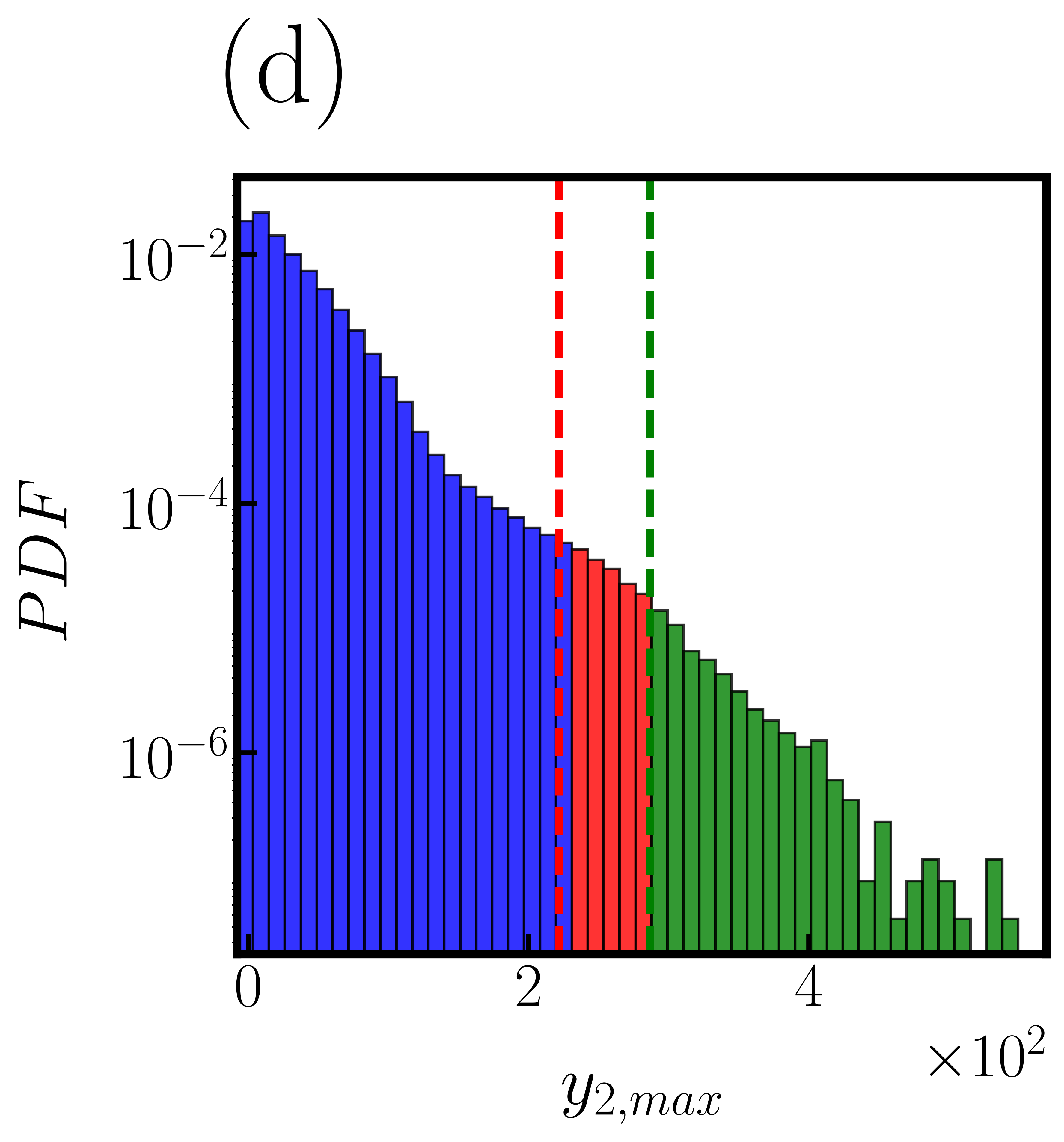}
    \caption{(a) \& (c) show the PDF distributions for the bounded chaotic velocity profiles of oscillators 1 \& 2, respectively, for interaction strength $\epsilon=5.15463$ and (b) \& (d) show the PDFs for the extreme events associated with velocity time series profiles of oscillators 1 \& 2, respectively,  for interaction strength value $\epsilon=13.1067$.}
    \label{pdf_two}
\end{figure}
Building on the preceding analysis, we now generalize the study to a multi-particle system by adding nonlinear repulsive interactions among the constituent Higgs oscillators as shown in the schematic diagram in the Fig. \ref{SD}.

The dynamics of the coupled Higgs oscillators is described by the Hamiltonian
\begin{equation}
H=\sum_{i=1}^{2}\left[
\frac{y_i^2}{2(1+\kappa x_i^2)^2}
+\frac{\omega_0^2x_i^2}{2}
\right],
\end{equation}
where the first term represents the nonlinear kinetic energy, while the second term denotes the harmonic potential contribution. The interaction between the oscillators is introduced through the coupling function
\begin{equation}
Q_i=\frac{x_i}{(1+\kappa x_i^2)^2}, \qquad i = 1, 2.
\end{equation}
Including the interaction, linear damping, and periodic forcing, the equations of motion for two-coupled Higgs oscillator takes the form
\begin{eqnarray}
\dot{x}_1 &=& y_1, \\
\dot{y}_1 &=& \frac{2\kappa x_1 y_1^2}{1+\kappa x_1^2}-\omega_0^2x_1(1+\kappa x_1^2)^2 -\epsilon\omega_0^2(Q_2-Q_1) \nonumber\\
&-&\alpha y_1 +f\cos(\omega_{e} t), \\ \nonumber \\
\dot{x}_2 &=& y_2,  \\
\dot{y}_2 &=& \frac{2\kappa x_2 y_2^2}{1+\kappa x_2^2}-\omega_0^2x_2(1+\kappa x_2^2)^2-\epsilon\omega_0^2(Q_1-Q_2) \nonumber \\
&-&\alpha y_2 +f\cos(\omega_{e} t),
\label{eqn-coupled-2}
\end{eqnarray}
where $\epsilon$ denotes the interaction strength, $\alpha$ is the linear damping coefficient, and $f\cos(\omega_{e} t)$ represents the external periodic forcing of amplitude $f$ and frequency $\omega_{e}$.
 Differentiating the Hamiltonian with respect to time yields the energy variation equation
\begin{equation}
\dot{H}_i=
\frac{y_i\dot{y}_i}{(1+\kappa x_i^2)^2}
-\frac{2\kappa x_i y_i^3}{(1+\kappa x_i^2)^3}
+\omega_0^2x_i y_i.
\end{equation}
Substituting the equations of motion into the above expression and simplifying, the intrinsic conservative contributions cancel exactly, leaving only the interaction, damping and forcing contributions, 
\begin{align}
\dot{H}_1 &=
-\epsilon\omega_0^2
\frac{y_1}{(1+\kappa x_1^2)^2}(Q_2-Q_1)
-\alpha\frac{y_1^2}{(1+\kappa x_1^2)^2} \nonumber \\
&+f\frac{y_1\cos(\omega_{e} t)}{(1+\kappa x_1^2)^2},  \\
\dot{H}_2 &=
-\epsilon\omega_0^2
\frac{y_2}{(1+\kappa x_2^2)^2}(Q_1-Q_2)
-\alpha\frac{y_2^2}{(1+\kappa x_2^2)^2} \nonumber \\
&+f\frac{y_2\cos(\omega_{e} t)}{(1+\kappa x_2^2)^2}.
\end{align}
Thus, the total rate of energy variation can be decomposed as
\begin{equation}
\dot{H}_i=
\dot{H}_i^{\mathrm{int}}
+\dot{H}_i^{\mathrm{damp}}
+\dot{H}_i^{\mathrm{force}},
\end{equation}
where the interaction term governs energy exchange between oscillators, the damping term accounts for irreversible energy dissipation and the forcing term is responsible for external energy injection into the system.

The resulting equations of motion governing the dynamics of the system are given by
\begin{eqnarray}
 \dot{x_{i}} &=& y_{i},\\
 \dot{y_{i}} &=& \dfrac{2\kappa x_{i}}{(1+\kappa x_{i}^2)}y_{i}^2 -\omega_0^2(1+\kappa x_{i}^2)^2 x_{i} \nonumber \\ 
 & & -\alpha y - \epsilon \Sigma_{j=1}^{N} \omega_0^2\left(Q_j- Q_i\right)\nonumber \\
  && + fcos(\omega_{e}t),\\
  \dot{H}_i &=&
-\epsilon\omega_0^2
\frac{y_i}{(1+\kappa x_i^2)^2}\Sigma_{j=1}^{N}(Q_j-Q_i)
-\alpha\frac{y_i^2}{(1+\kappa x_i^2)^2} \nonumber \\
&+&f\frac{y_i\cos(\omega_{e} t)}{(1+\kappa x_i^2)^2}~~~~(i=1,2,...,N).
 \label{eqn-coupled-n}
\end{eqnarray}
In the following analysis, we first consider the standard case of two coupled oscillators (N=2) to elucidate the underlying interaction dynamics. We then extend the analysis to a larger network with N=10, chosen as a representative example to demonstrate the collective behavior. The choice of N=10 is illustrative and does not imply that the analysis is restricted to this network size only.

For numerical integration, we employ the Runge - Kutta fourth order method with a fixed step size of $h = 0.01$. In particular, we investigate the dynamics of the system by tuning the system parameters and explore the mechanism underlying the genesis of extreme events in the coupled Higgs oscillator system.

\begin{figure}[h!]
    \centering
    \includegraphics[width=1.0\linewidth]{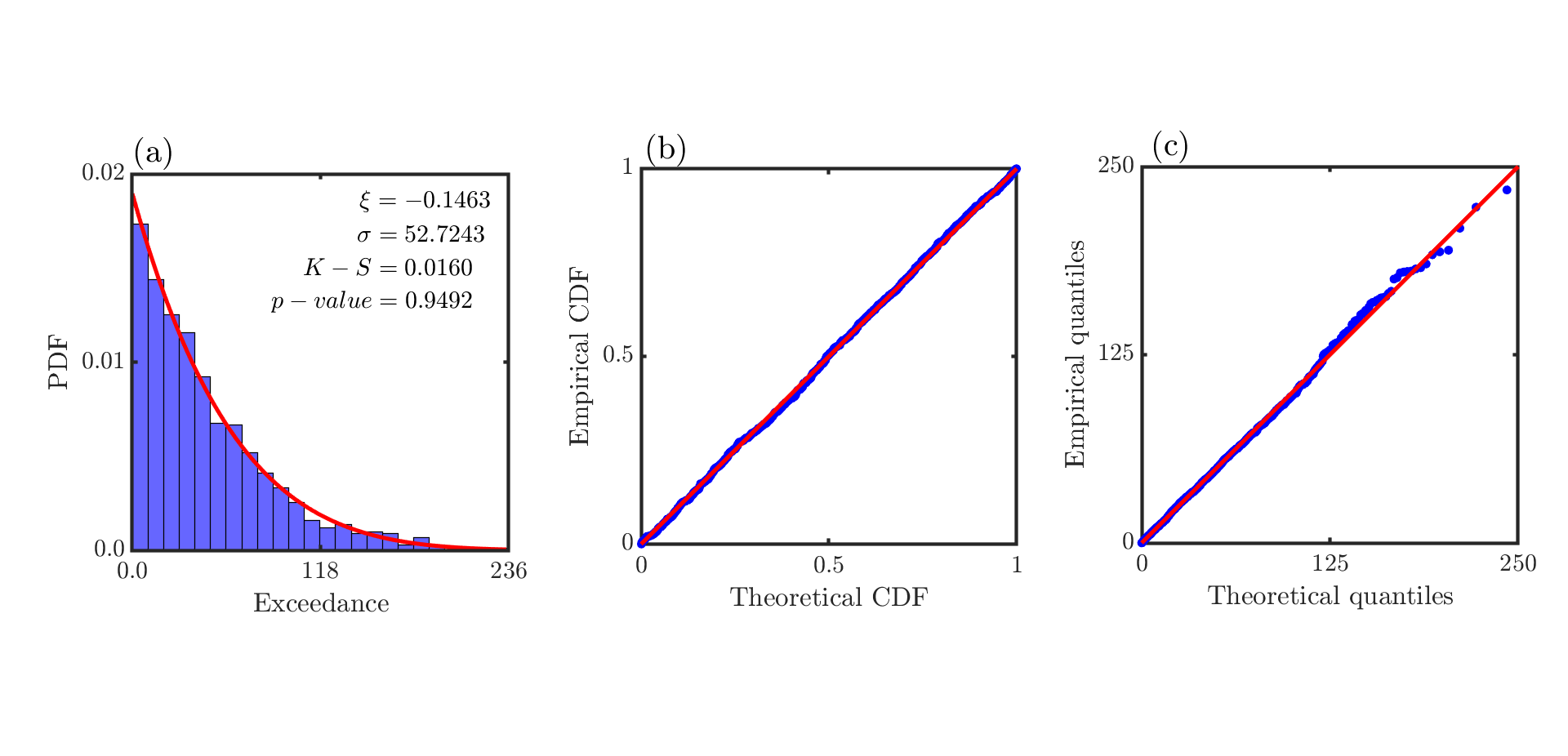}
    \includegraphics[width=1.0\linewidth]{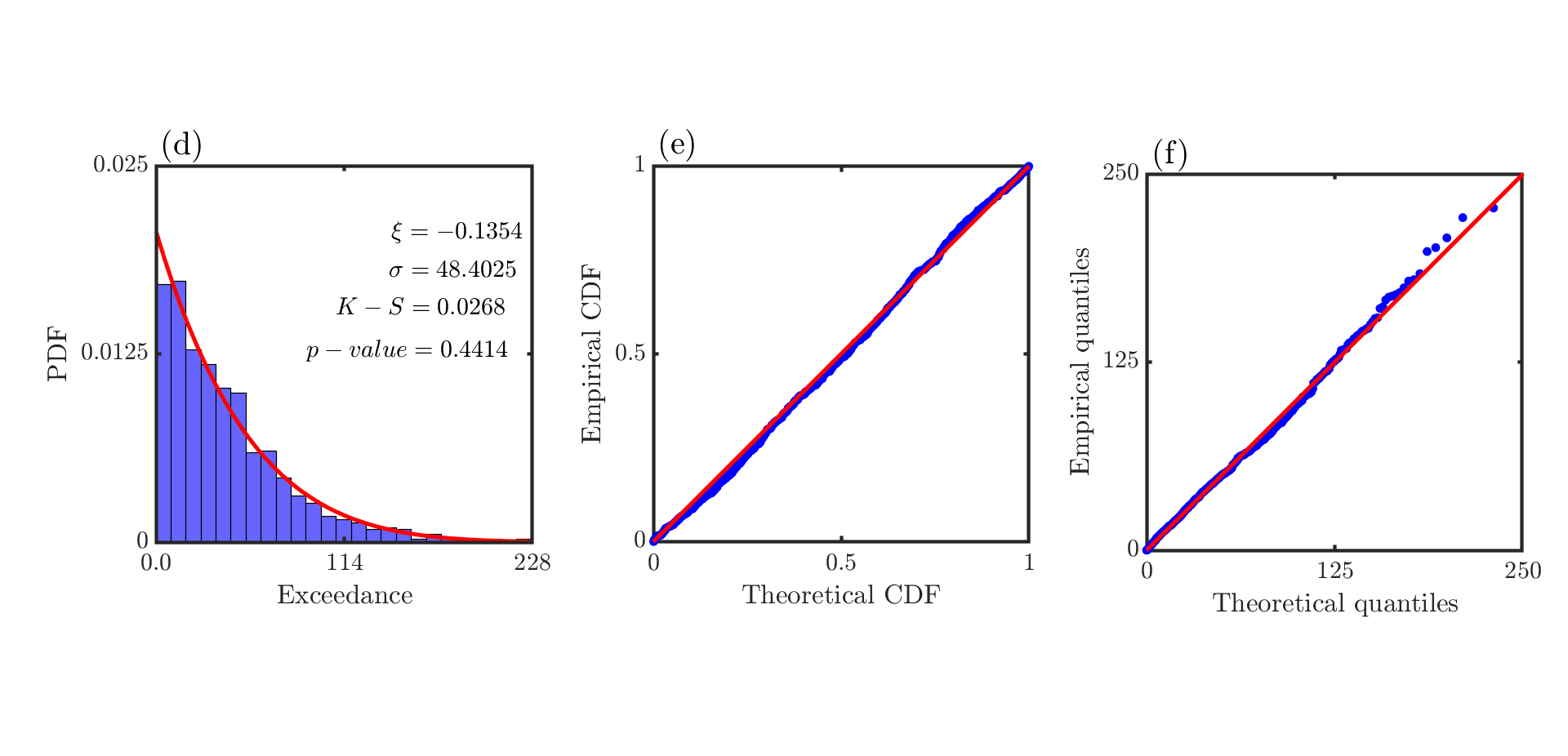}
    \caption{(a) \& (d) show the exceedance distributions with Generalized Pareto distribution (GPD) fit in red line for the extreme events associated with velocity time series profiles of oscillators 1 \& 2, respectively,  for interaction strength value $\epsilon=13.1067$. Figs(b)-(e) \& (c)-(f) show the Q-Q and P-P plots  which validate the tail behaviour, goodness of the fit and the chosen EE threshold. }
    \label{GPD}
\end{figure}

\section{Two coupled Higgs oscillators}

We consider a system of two Higgs oscillators, coupled via the mass–interaction term as specified in Eq.(\ref{eqn-coupled-n}), with oscillator 1 represented by the coordinates $(x_1, y_1)$ and oscillator 2 by $(x_2, y_2)$. The initial conditions (ICs) are arbitrarily generated from the random number generator and the values are ($x_{1}(0),y_{1}(0),x_{2}(0),y_{2}(0))= (0.32597,0.54905,0.21858,0.89424)$. These initial conditions are fixed throughout the analysis and the dynamics is observed after a sufficient transient time. When $\kappa ~,\alpha ,f $ and $\epsilon$ tend to $0$, the system behaves as decoupled linear harmonic oscillators. A slight increase in the value of $\kappa$ leads to the emergence of an isolated non-trivial periodic attractor. By gradually increasing the interaction strength parameter $\epsilon$, we observe a sequence of attractors such as periodic, torus and chaotic attractors.

Exploration of the parameter space revealed that the position variable does not exhibit extreme-event-like behaviour in the considered parameter regimes. In contrast, extreme events are observed in the velocity variable (y). Hence, through the statistical analysis of the velocity dynamics, we examine the occurrence of extreme velocity fluctuations in the Higgs oscillator (see Fig.~\ref{TSPWL}). To investigate the mechanism responsible for the emergence of such slow and extreme velocity dynamics in the system, we first perform a bifurcation analysis to identify the dynamical route through which extreme events emerge in the two coupled Higgs oscillator system. We then choose suitable control parameter for bifurcation analysis by fixing all the other system parameters as constant. After analysing the system in the presence of parametric perturbations, we consider the coupling strength between the oscillators, $\epsilon$, as the control parameter, while keeping  $\kappa=0.21$, $\omega_{0}^{2} = 0.1$, $\alpha = 0.00595$, $f=4.3251$ and $\omega_{e} = 0.21$ fixed throughout the study. The control parameter is varied over the region $\epsilon \in (0.0,15.0)$. Also, we calculate the extreme event threshold (say $H_{th}$) to keep track of the parameter that generates the onset of extreme events. Generally, in the literature, the emergence of EEs is characterized by a threshold value \cite{chowdhury2022extreme}, which can be calculated by using the formula, 
\begin{eqnarray*}
    H_{th}= \langle y_{n} \rangle + \gamma*\sigma(y_{n}), \label{hthreshold}
\end{eqnarray*} 
where  $\langle y_{n} \rangle$ is the time average of the peaks and $\sigma(y_{n})$ is the standard deviation and $\gamma$ is the threshold qualifier parameter which controls the degree of rareness of the EEs. Typically, $\gamma$ runs from $4$ to $8$. Accordingly, the events are characterized into extreme and super extreme events based on this $\gamma$ parameter \cite{PhysRevE.96.012216, vijay2023superextreme}. Throughout our analysis, we fix $\gamma= 6$ for identifying EEs and $\gamma= 8$ to observe SEEs. However, in the bifurcation analysis, only the EE threshold ($\gamma=6$) is shown, which is indicated by the black and green dotted lines for oscillator-1 and oscillator-2, respectively. Subsequently, it will be shown that the chosen EE threshold is appropriate for capturing the tail behaviour. Upon scanning the parameter $\epsilon$ over the interval $\epsilon \in (0.0,1.28674)$, the attractor remains quasiperiodic for both the oscillators as can be seen in Fig. \ref{bif-lya}(a) above characterized by $\Lambda_{max} \approx 0.0$ which is shown (see also the inset plot in Fig. \ref{bif-lya}(b)). After $\epsilon = 1.28674$, the system transits to chaos where $\Lambda_{max}$ becomes positive. Although, the oscillators $1$ and $2$ transit to chaos,  they remain bounded under the extreme event threshold. The system remains bounded and chaotic until $\epsilon = 8.6739$. On further increasing the value of $\epsilon \in (8.6739,15.0)$ the attractors suddenly expand traversing the extreme event threshold creating a sparser set of blue and red dots  which corresponds to local maxima of oscillator-$1$ \& oscillator-$2$ crossing the defined EE threshold in the bifurcation diagram in Fig. \ref{bif-lya}(a). Also, we note that the values of $\Lambda_{max}$  slowly increase on increasing the strength of $\epsilon \in (1.28674,15.0)$ which clearly indicates that there is greater divergence property of the attractor during the EEs. Furthermore, in our previous studies on the damped-driven Higgs oscillator, the extreme events occurred via period doubling followed by an interior crisis \cite{wasif2025extreme}. On the other hand, in the present study of the two coupled Higgs oscillators, we observe quasi-periodic route leading to chaos and subsequently to EEs and this route adds additional novelty to the present work.

\subsection{Time series and the observation of extreme events}

From the bifurcation diagram (Fig. \ref{bif-lya}), we have identified the parameters that generate extreme events. To further characterize the underlying dynamics in these regimes, the time series of position, velocity and the instantaneous Hamiltonian (energy) of the oscillators - $1\& 2$ at  $\epsilon=5.15463$  are presented in Figs. \ref{TSPWL}(a)(i-iii) and \ref{TSPWL}(d)(i-iii). We observe that the time trajectories of all the three state variables, $x$, $y$, and $H$, remain bounded under the adopted EE threshold. The calculated EE and SEE threshold values for all the state variables of both the oscillators are listed in Table~\ref{tab:threshold}. Furthermore, for the EE case with $\epsilon = 13.1067$ , we find that while the position variable does not satisfy the adopted extreme-event threshold, the velocity and instantaneous energy variables exceed both the EE and SEE thresholds, $H_{th}$ and $H_{ths}$, respectively, as shown in Figs. \ref{TSPWL}(b)(i-iii) and  \ref{TSPWL}(d)(i-iii). It is worth noting that, over a longer integration time of $5\times10^{7}$ time units, the instantaneous energy exhibits a slight upward trend accompanied by fluctuations. Consequently, the corresponding energy time series does not satisfy the adopted EE threshold criterion. In contrast, the velocity variable $y$ does not exhibit any noticeable trend over the entire integration time and continues to satisfy the adopted EE threshold criterion. So, in the subsequent section we employ the velocity variable for further characterization of the dynamics and to understand the emergence of fast fluctuations in the coupled Higgs oscillators.

Furthermore, we analyze the influence of the mass function on the velocities via Figs. \ref{TSPWL}(c) \& \ref{TSPWL}(f), which illustrate the power-law relations between the conformal mass factor $m(x)=\frac{1}{(1+\kappa x^{2})^{2}}$ and the absolute maximum velocities \bigg( $|y^{1,2}_{max, BC}|$ vs $m(x_{n})$ for BC and $|y^{1,2}_{max, EE}|$ vs $m(x_{n})$ for EE \bigg) along with their corresponding power-law fits for oscillators 1 \& 2, respectively, denoted as $y^{1,2~(fit)}_{max, BC}$ and $y^{1,2~(fit)}_{max, EE}$.
(\noindent\textbf{Note on notation:}
We emphasize that the function $m(\cdot)$ is defined on the state variable $x$ and should be interpreted carefully depending on the argument. In particular, (i) the function evaluated at a generic state variable $x$ is denoted by $m(x)$, (ii) the time-dependent evaluation of the function along the system trajectory $x(t)$ is represented by $m(x(t))$, 
(iii) the function evaluated at discrete points $x_n$, where $x_n$ represents the locations associated with the local maxima $|y_{\max}|$, is denoted by $m(x_n)$. These three notations should not be used interchangeably, as they are fundamentally different. Also, the notation $z$ is used to denote the space of $m(x_{n})$ where $|y_{max}|$ occurs.)

\subsection{Statistical Analysis of Extreme Events}

After observing the occurrence of EEs in the time series, we now analyze the statistics for local maxima values of  bounded chaos and EEs by presenting the  probability density function (PDF) for the oscillators 1 \& 2. For bounded chaos, the PDF shows no distribution of velocity maxima beyond the threshold ($H_{th}$), as depicted in Figs. \ref{pdf_two}(a) and  \ref{pdf_two}(c) whereas for EEs, the PDF shows the distribution of velocity maxima beyond the threshold ($H_{th}$) as in Figs. \ref{pdf_two}(b) and \ref{pdf_two}(d). Furthermore, we analyze the tail behaviour of the extreme event distributions by employing the peak-over-threshold framework together with the Generalized Pareto distribution (GPD), which enables efficient characterization of the tail behavior and provides reliable estimates even when the available data are limited \cite{lucarini2016extremes, albeverio2006extreme}. The general equation for the GPD function is  given by $F(h|k,\sigma) = \frac{1}{\sigma}(1 + \frac{kh}{\sigma})^{-1-\frac{1}{k}} $ when $k \neq 0$ and $F(h|k,\sigma) = \frac{1}{\sigma}e^{-\frac{h}{\sigma}}$ for $k=0$. 
Here, $\sigma$ is the scale parameter, $k$ is the shape parameter and $\{h_n\}$ denotes the exceedance values, with $h_n = EE_n - H_{\mathrm{th}}$, where $\{EE_n\}$ denotes the set of peak values exceeding the EE threshold $H_{\mathrm{th}}$. We then plotted the distribution of exceedances and fitted the GPD curve for oscillator $1$ and $2$ in Figs. \ref{GPD}(a) and \ref{GPD}(d) using MATLAB GPD fitter with the estimated scale parameter ($\sigma$) and shape parameter ($k$) for oscillator $1$, given by $ k_{1} = -0.1463$, $\sigma_{1} = 52.7243$ and for oscillator $2$, given by $ k_{2} = -0.1354$, $\sigma_{2}=48.4025$. Since the sign of the parameter $k$ is negative for both the oscillators, the corresponding GPDs possess a finite upper bounds, indicating a bounded tail in the exceedance distributions.

\begin{table}[h]
\footnotesize
\centering
\renewcommand{\arraystretch}{1.30}
\begin{tabular}{|c|c|c|c|}
\hline
\textbf{Parameter} & \textbf{$x$} & \textbf{$y$} & \textbf{$H$} \\
\hline

\multirow{2}{*}{$\epsilon=5.15463$}
&
$\begin{aligned}
H_{\mathrm{th},1} &= 34.205\\
H_{\mathrm{ths},1}&=44.151
\end{aligned}$
&
$\begin{aligned}
H_{\mathrm{th},1} &=127.310\\
H_{\mathrm{ths},1}&=162.915
\end{aligned}$
&
$\begin{aligned}
H_{\mathrm{th},1} &=19.348\\
H_{\mathrm{ths},1}&=24.0554
\end{aligned}$
\\
\cline{2-4}

&
$\begin{aligned}
H_{\mathrm{th},2} &=34.157\\
H_{\mathrm{ths},2}&=44.1146
\end{aligned}$
&
$\begin{aligned}
H_{\mathrm{th},2} &=127.441\\
H_{\mathrm{ths},2}&=163.146
\end{aligned}$
&
$\begin{aligned}
H_{\mathrm{th},2} &=19.3489\\
H_{\mathrm{ths},2}&=24.0580
\end{aligned}$
\\
\hline

\multirow{2}{*}{$\epsilon=13.1067$}
&
$\begin{aligned}
H_{\mathrm{th},1} &=38.781\\
H_{\mathrm{ths},1}&=49.943
\end{aligned}$
&
$\begin{aligned}
H_{\mathrm{th},1} &=220.477\\
H_{\mathrm{ths},1}&=284.900
\end{aligned}$
&
$\begin{aligned}
H_{\mathrm{th},1} &=28.6475\\
H_{\mathrm{ths},1}&=35.766
\end{aligned}$
\\
\cline{2-4}

&
$\begin{aligned}
H_{\mathrm{th},2} &=38.536\\
H_{\mathrm{ths},2}&=50.013
\end{aligned}$
&
$\begin{aligned}
H_{\mathrm{th},2} &=225.124\\
H_{\mathrm{ths},2}&=291.004
\end{aligned}$
&
$\begin{aligned}
H_{\mathrm{th},2} &=29.0339\\
H_{\mathrm{ths},2}&=36.2488
\end{aligned}$
\\
\hline

\end{tabular}

\caption{Threshold values ($H_{th}$ for EEs and $H_{ths}$ for SEEs) corresponding to the first and second oscillators for the position ($x$), velocity ($y$), and energy ($H$) variables at two representative coupling strengths.}
\label{tab:threshold}
\end{table}

Additionally, the goodness of the GPD fit is assessed using both probability-probability (P–P) and quantile-quantile (Q–Q) plots. The P–P plot is constructed by comparing the empirical cumulative distribution function (ECDF) of the exceedance data with the theoretical cumulative distribution function (TCDF) of the fitted GPD, while the Q–Q plot compares the empirical and theoretical quantiles. Furthermore, the Kolmogorov–Smirnov (KS) goodness-of-fit test is performed to examine whether the exceedances of oscillators $1$ and $2$ are consistent with the fitted GPD. The corresponding KS statistic values are found to be 0.0160 and 0.00268 for  oscillators 1 and 2, respectively, indicating excellent agreement between the empirical and theoretical cumulative distributions. The associated $p-values$ are 0.9492 for oscillator 1 and 0.4414 for oscillator 2, both of which exceed the conventional $p-value$ threshold of 0.05 used in the Kolmogorov–Smirnov goodness-of-fit test. Hence, the KS test fails to reject the null hypothesis that the exceedance data follow a Generalized Pareto Distribution. Together with the close agreement observed in the P–P and Q–Q plots, these results indicate that the GPD provides an appropriate statistical model for the exceedance data. Consequently, the results support the choice of $\gamma=6.0$ as an appropriate EE threshold parameter for distinguishing extreme events (EEs) from non-extreme events (NEEs).

\subsection{Effect of external forcing ($f$)}
 
To investigate the role of external forcing in the genesis of extreme events (EEs), we construct a two-parameter phase diagram that identifies the regions corresponding to non-extreme events (NEEs-grey), extreme events (EEs-yellow), and super extreme events (SEEs-red). In addition, we quantify the energy exchange between the coupled oscillators through the time-averaged synchronization error during the time interval (${T_{i}} $ and ${T_{f}}$) in their instantaneous energy, defined as

\begin{equation} \langle \Delta H \rangle = \frac{1}{T} \int_{T_{i}}^{T_{f}} \left|H_1(t)-H_2(t)\right|\,dt, \label{eq:sync_error_x} \end{equation}

where $H_1(t)$ and $H_2(t)$ denote the instantaneous energy of oscillators 1 and 2, respectively. Scanning the parameter space $(f,\epsilon)\in(0.0,15.0)\times(0.0,8.0)$ while keeping all other parameters fixed, we obtain the phase diagrams shown in Fig.~\ref{two_phase}. For small coupling strengths $\epsilon$, the system remains entirely below the statistical threshold over the whole range of the forcing amplitude $\epsilon$, indicating the absence of extreme events [Fig.~\ref{two_phase}(a)]. In this regime, the time-averaged synchronization error is nearly zero [Fig.~\ref{two_phase}(b)], implying that both the oscillators exchange energy symmetrically and therefore maintain almost identical instantaneous energy (More details of Figs.~\ref{two_phase}(b) and \ref{two_phase}(d) regarding the synchronization error in energy is provided in the Appendix).

As the forcing amplitude exceeds the value $f\approx2.0$ and the coupling strength increases beyond $\epsilon\approx2.5$, the occurrence of both EEs and SEEs becomes increasingly prominent. Although isolated islands of nearly zero synchronization error persist within the region $\epsilon\in(2.0,15.0)$, these correspond predominantly to NEE dynamics. Away from these islands, the synchronization error increases progressively, as indicated by the radial color gradient in Fig.~\ref{two_phase}(b). This demonstrates that the emergence of extreme events is closely associated with an increasing asymmetry in the energy distribution between the two oscillators. Consequently, the unequal exchange of energy destabilizes the coupled dynamics, facilitating the occurrence of large velocity excursions that manifest as extreme and super extreme events.

\subsection{Effect of curvature constant ($\kappa$)}
\begin{figure}[]
    \centering
      \includegraphics[width=0.47\linewidth]{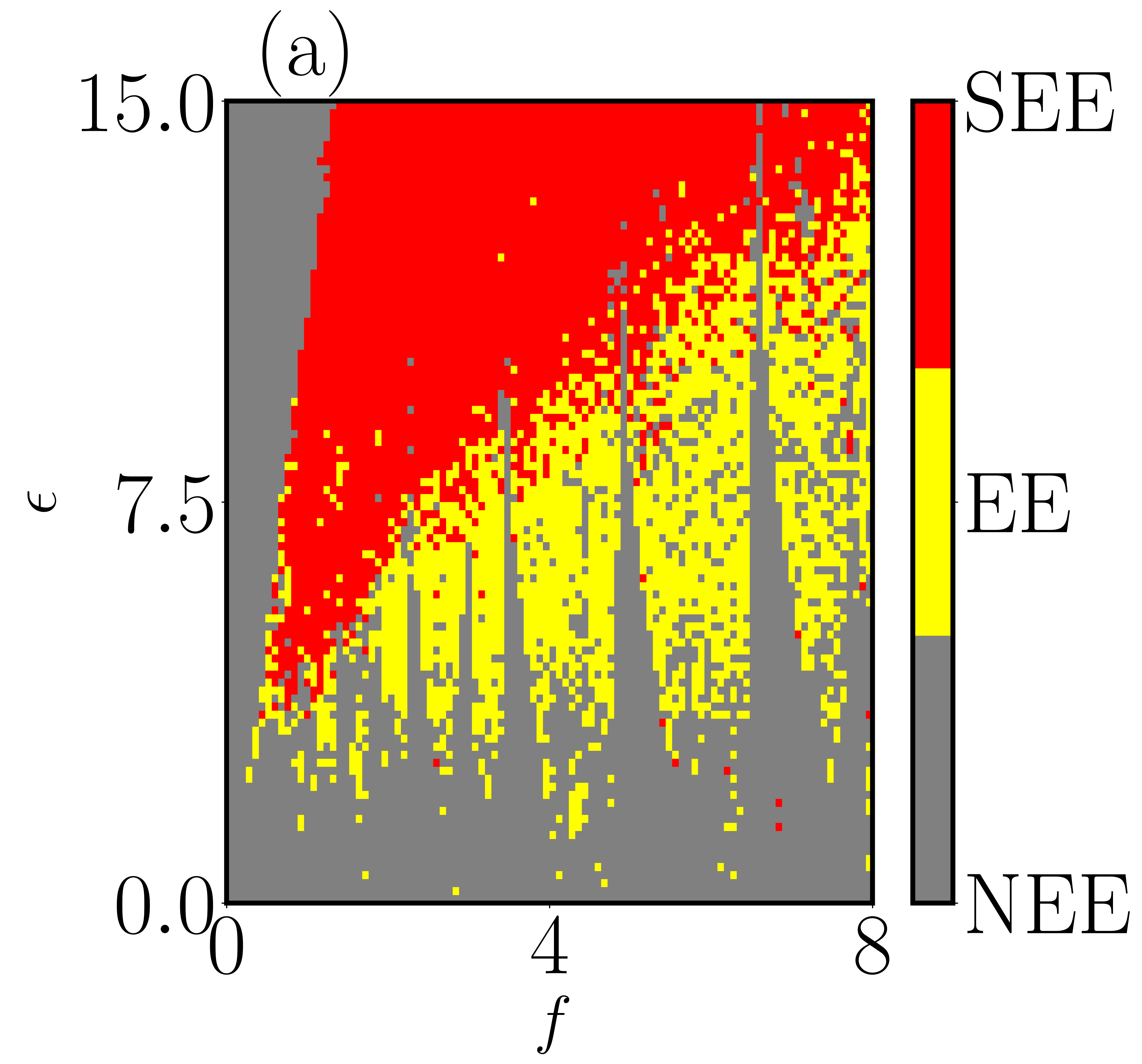}
      \includegraphics[width=0.47\linewidth]{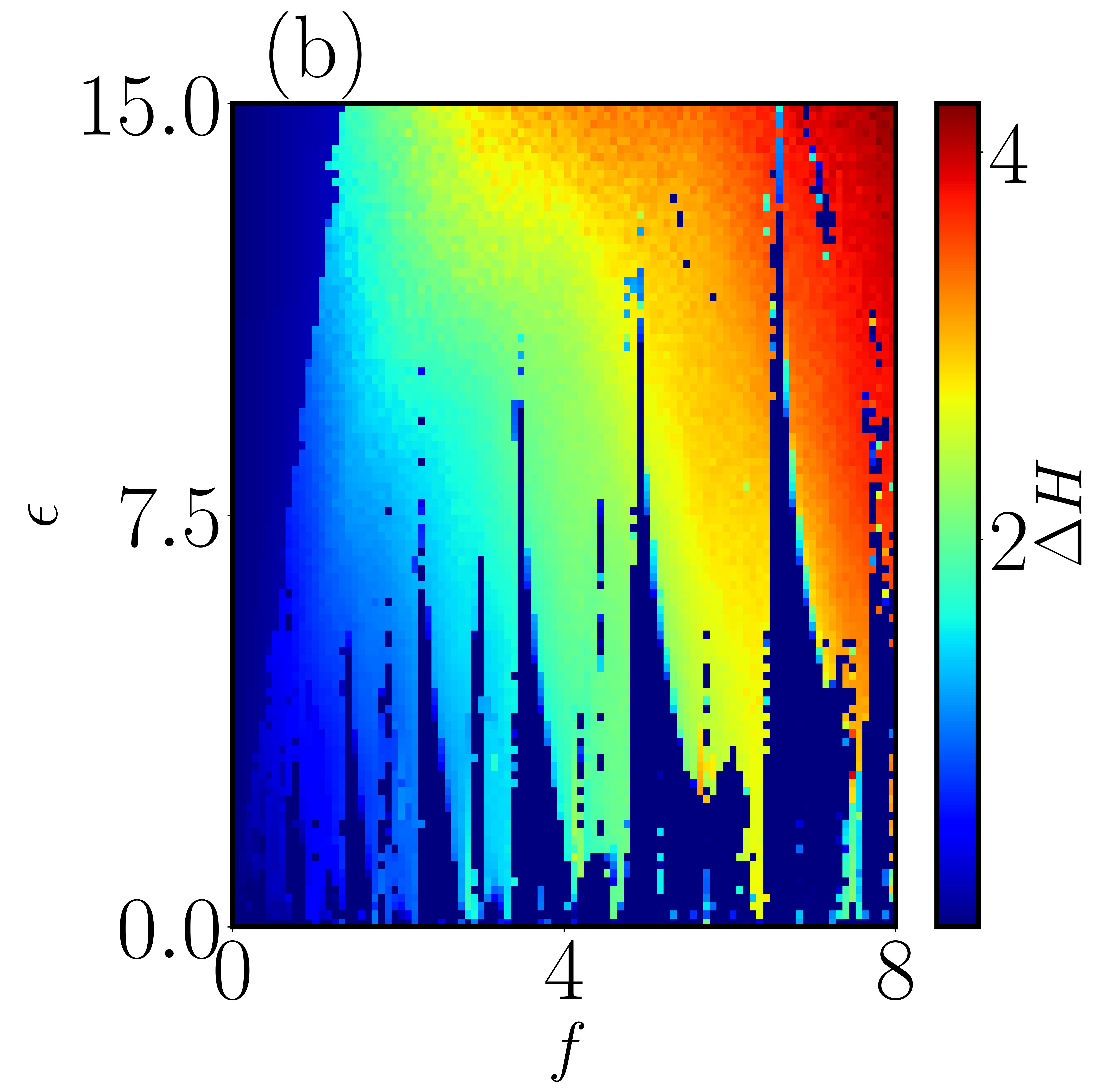}
      \includegraphics[width=0.47\linewidth]{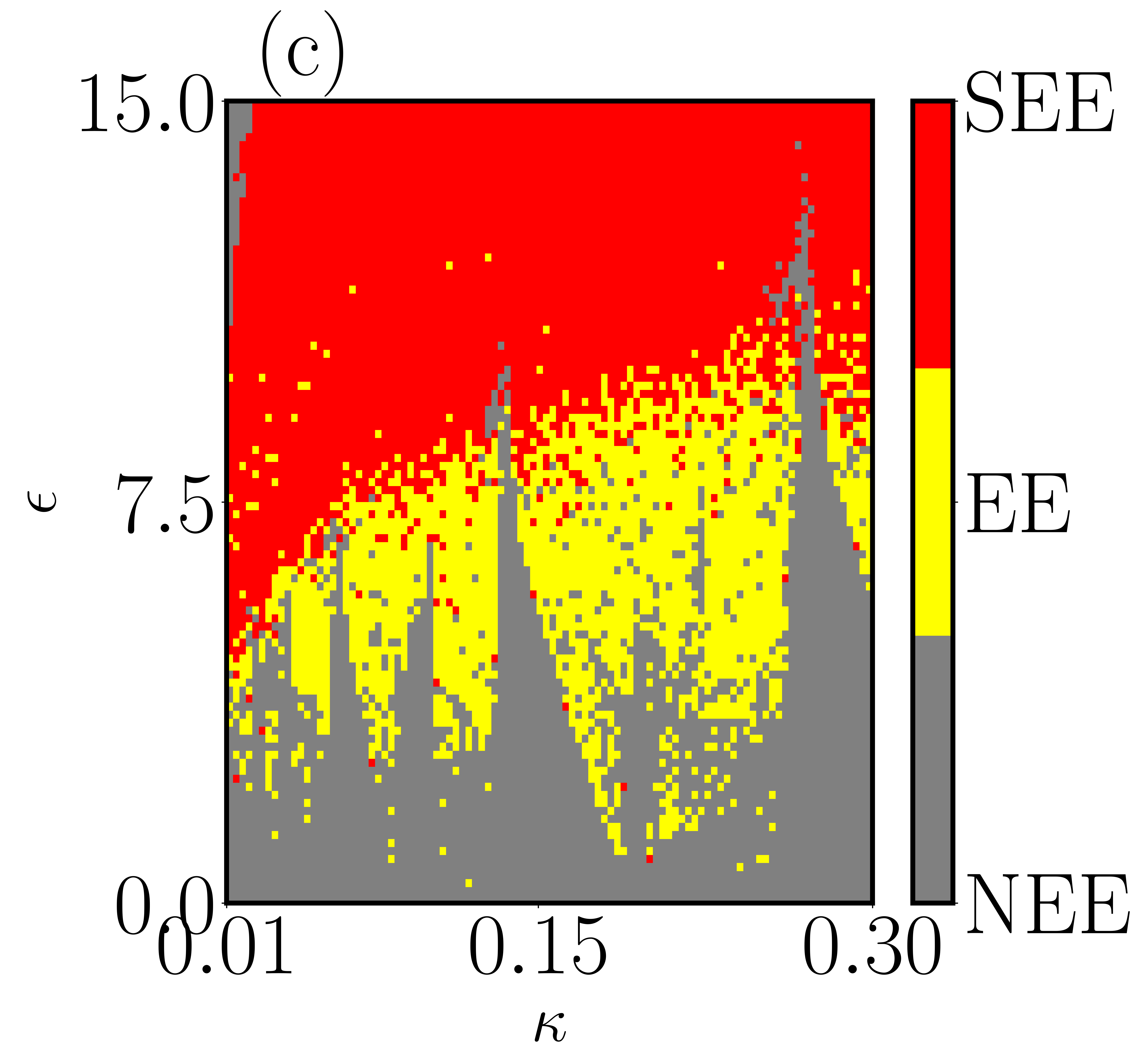}
      \includegraphics[width=0.47\linewidth]{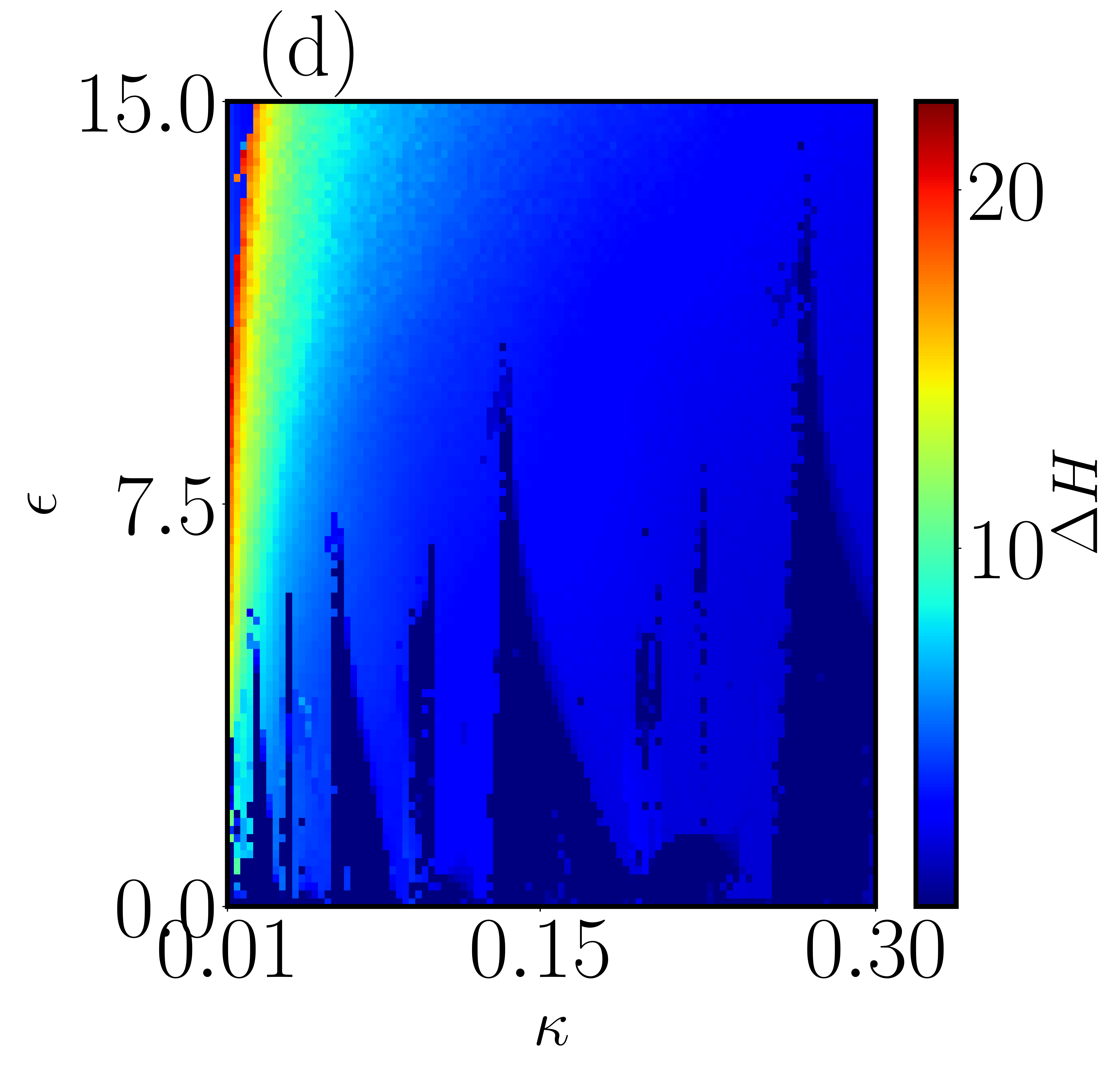}
\caption{(a)-(c) Two parameter phase diagram for ($f\in (0.0,15.0)$ vs $\epsilon  \in (0.0,8.0) $) \& ($\kappa \in (0.01,0.30)$ vs $\epsilon \in (0.0,15.0)$) delineating the regions of non-extreme event (NEE)-grey, extreme event (EE)-yellow and super extreme event region (SEE)-red. Here (b) \& (d) are the corresponding time averaged synchronization error between the instantaneous energies for two-coupled Higgs oscillators.}
    \label{two_phase}
\end{figure}
In this subsection, we investigate the combined influence of the interaction strength ($\epsilon$) and the curvature parameter ($\kappa$) on the emergence of extreme events by scanning the parameter space $\epsilon\in(0.0,15.0)$ and $\kappa\in(0.01,0.3)$, while keeping all other system parameters fixed. A detailed examination of the phase diagram reveals that the system remains in the non-extreme-event (NEE) regime for weak interaction strengths. As the interaction strength is increased beyond $\epsilon\approx2.75$, regions corresponding to extreme events (EEs) and super extreme events (SEEs) progressively emerge over a broad range of curvature values, as illustrated in Fig.~\ref{two_phase}(c). The corresponding time-averaged energy synchronization error, shown in Fig.~\ref{two_phase}(d), indicates that the NEE regime is characterized by isolated islands with nearly zero synchronization error, reflecting an almost symmetric distribution of energy between the oscillators. In contrast, the EEs and SEEs are accompanied by a finite synchronization error, signifying an increasingly asymmetric redistribution of energy between the coupled oscillators.

For interaction strengths exceeding $(\epsilon \approx 2.75)$, the energy synchronization error increases progressively as the curvature parameter $\kappa$ decreases. In particular, the synchronization error becomes significantly larger in the low-$\kappa$ regime, suggesting an increasing asymmetry in the energy exchange between the coupled oscillators [see Figs.~\ref{two_phase}(c) and \ref{two_phase}(d)]. These observations demonstrate that the interaction strength acts as the primary control parameter governing the onset of extreme events, whereas the curvature parameter modulates the extent and persistence of the energy imbalance between the oscillators. The increasing synchronization error therefore provides clear evidence that the emergence of extreme events is closely associated with the asymmetric distribution of energy within the coupled system.

\subsection{Mechanism}

Extreme events in the dynamics of a single oscillator arise from instabilities in the phase space, including the presence of saddle points, singularities and saddle orbits. In most cases, EEs often emerge due to the interior crisis where the trajectory meets the boundaries of the stable and unstable manifolds which makes the trajectory to move along the boundary resulting in long excursions. These critical boundary points are referred to as crisis points. The other major mechanism is the intermittency crisis. At particular system parameters, periodic orbits are intercepted by chaotic bursts near a point where the collision of the stable and unstable periodic orbits occurs, leading to the saddle orbit. This saddle orbit mediates the sudden onset of extreme events.
\begin{figure}[h!]
    \centering
    \includegraphics[width=0.45\linewidth]{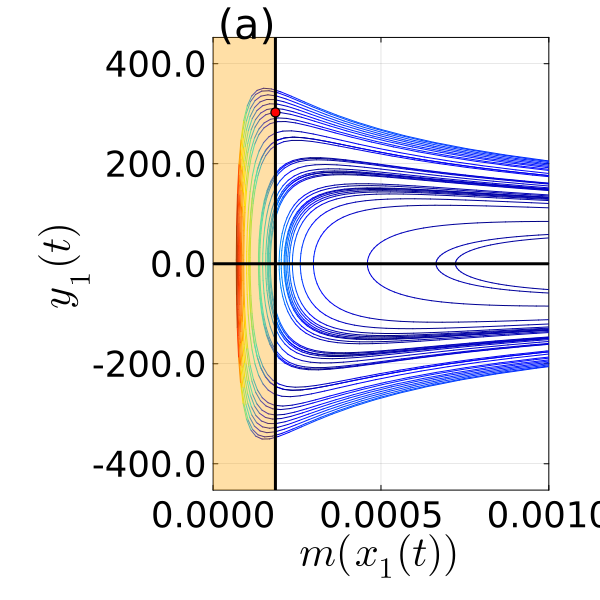}
    \includegraphics[width=0.45\linewidth]{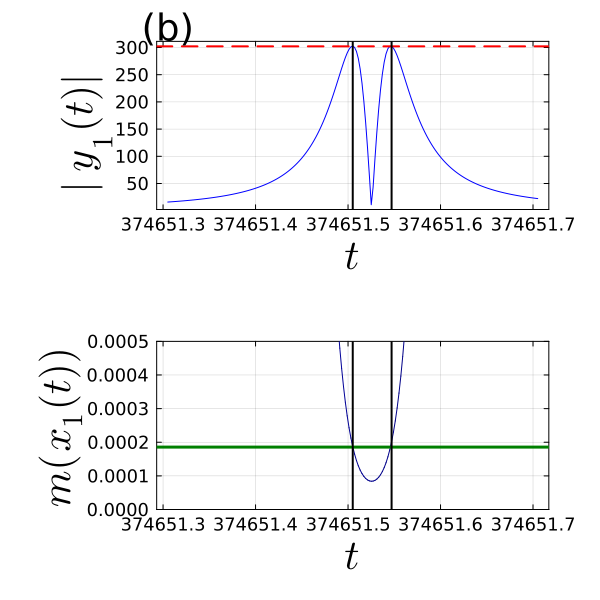}
    \includegraphics[width=0.45\linewidth]{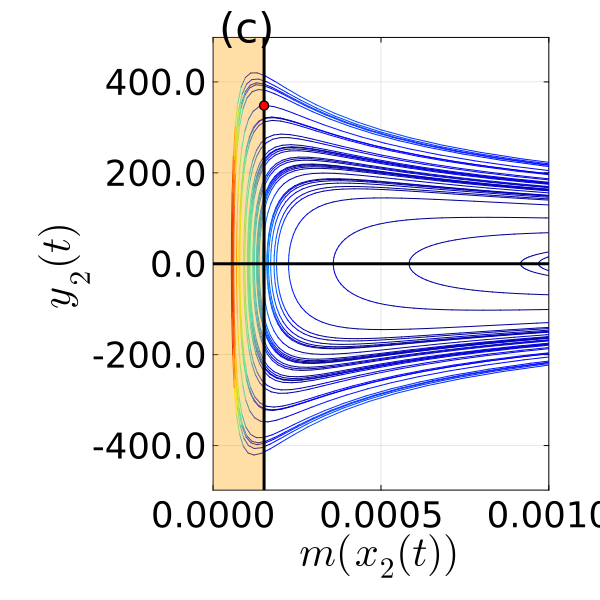}
    \includegraphics[width=0.45\linewidth]{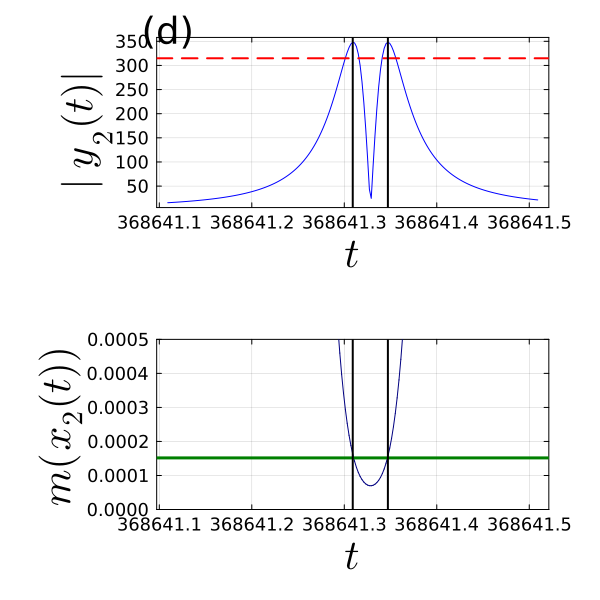}
     \caption{(a) and (c) show the zoomed (m(x(t))) vs (y(t)) plots for the EE chaotic velocity profiles of oscillators 1 and 2 at coupling strength ($\epsilon = 13.1067 $). (b) and (d) compare the velocity magnitude and mass profiles, where the green bisector lines indicate the turning-point masses of oscillators 1 and 2, obtained from the respective power-law relations.
}
    \label{mass_velocity_phase_space}
\end{figure}
In the coupled oscillators case, the genesis of extreme events arises from the occasional in-phase synchronization between the oscillators. On-off intermittency plays a crucial role where the trajectory of any one of the coupled oscillators traverses perpendicularly from the synchronization manifold, giving rise to the EEs. In this synchronization approach, the  mechanism of EEs can be determined by calculating the synchronization error between the oscillators. Many of these mechanisms were explored in certain ecological, neuronal and in some mechanical models. In this manuscript, we explore the genesis of EEs in the coupled non-polynomial position-dependent mass system. In our previous study, we analyzed EEs in a one-dimensional damped and forced Higgs oscillator by identifying phase space instabilities using Poincar\'{e} maps.

For understanding the EEs dynamics, one often visualizes the expansion of the attractor in the phase space and then locates the instabilities. By doing so, one hardly finds the relation between the phase space variables for the emergence of EEs. Therefore, we essentially visualize the shape of the attractor in $y(t)$ and $m(x(t))$ space. The connection between the maximum absolute velocity $|y_{max}|$ and the conformal mass $m(x)$ can be established through the power law relation given by $|y_{max}|\sim az^{b}$, where $a$ and $b$ represent the scaling exponent and power index. From this relations,  we can find the critical conformal mass $m^{*}(x)$ for extreme and bounded chaotic behaviour.

To obtain this relation, we collect the maxima of the velocity profile and their corresponding positions $x_{n}$. By doing so, we find certain inverse relation between the conformal mass factor $z$ vs $|y_{max}|$, where $z = m(x_n)$. This indicates that as $z$ approaches zero value (critical minimum mass), the velocity abruptly attains its maximum value $|y_{max}|$. To express the above statement in mathematical terms, we have plotted the power-law relation between $|y_{max}|$ and $z$ for the EEs occurring in oscillators 1 and 2 separately. The reason for taking  sets of EEs for establishing the power-law relation is that they span the maximum region in the phase space and hence can be generalized for bounded chaotic and other periodic sets. 

\begin{figure}[h!]
    \centering
      \includegraphics[width=0.95\linewidth]{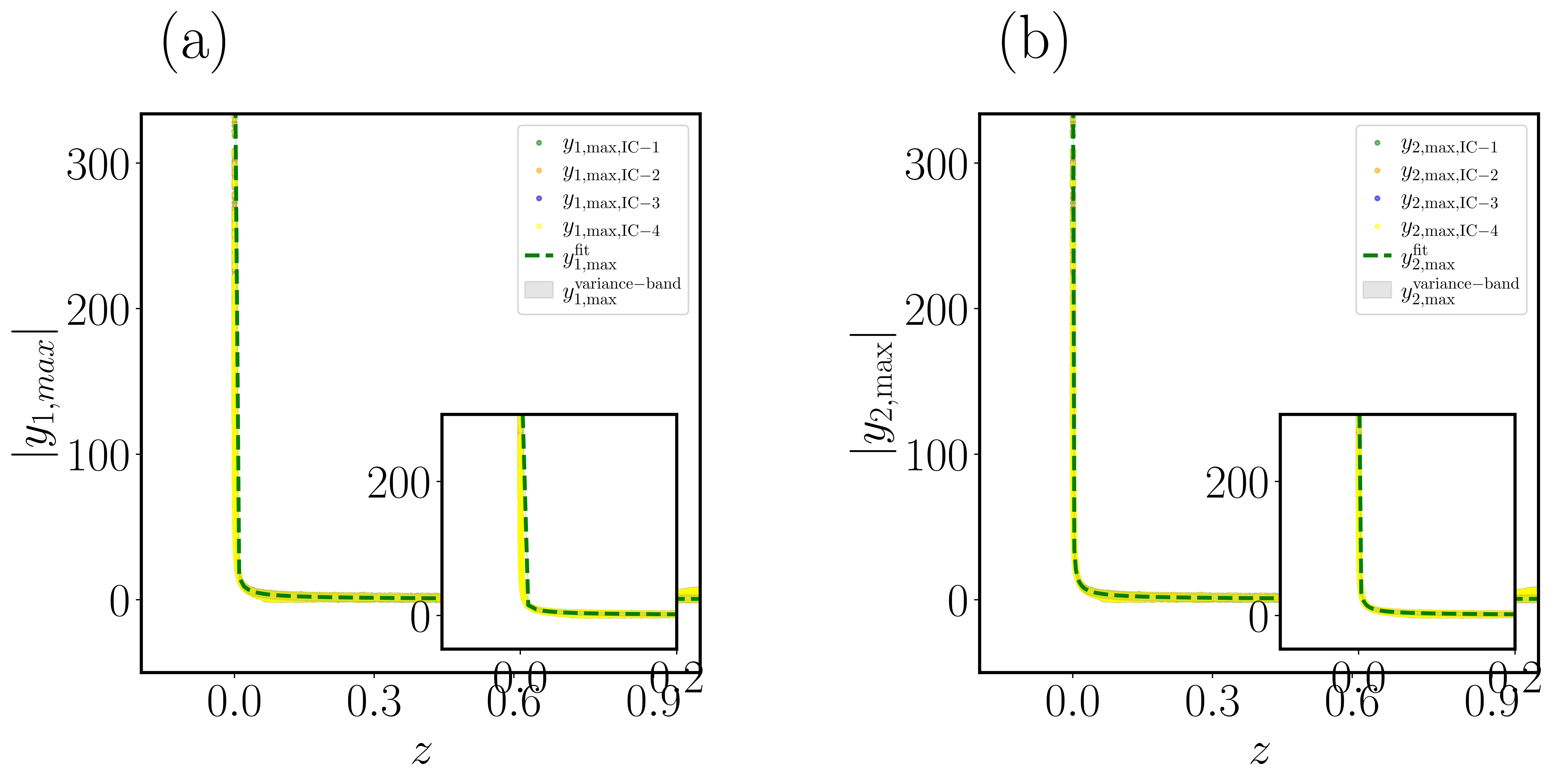}
       \caption{(a) \& (b) show the power law  fit in green dashed line for the extreme events associated with velocity time series profiles of oscillators 1 \& 2, respectively,  for interaction strength value $\epsilon=13.1067$.The $|y_{max}|$ vs $z$ are plotted for 4 different initial conditions(ICs) and ICs are taken in the form $IC_{1}=[x_{1}(0),y_{1}(0),x_{2}(0),y_{2}(0)]$ (green),
$IC_{2}=[-x_{1}(0),y_{1}(0),x_{2}(0),y_{2}(0)]$ (orange), 
$IC_{3}=[-x_{1}(0),-y_{1}(0),x_{2}(0),y_{2}(0)]$ (blue),
and $IC_{4}=[x_{1}(0),-y_{1}(0),x_{2}(0),y_{2}(0)]$ (yellow).}
    \label{powerICs}
\end{figure}

We have shown the power-law relations in Figs.~\ref{TSPWL} (c) and \ref{TSPWL}(f) where we have plotted $|y_{max}|$ vs $z$ relation for bounded chaotic (BC) attractor (green dots) and extreme event (EE) attractor (red dots) for oscillator 1  and oscillator 2. The corresponding scaling power law fit for bounded chaos and EEs for oscillators 1  $\&$ 2 are represented in yellow and blue dashed lines. The scaling exponents of oscillator 1 for bounded chaos are $a=0.4556$ and $b=-0.7602$ and their values for EEs are $a=0.5005$ and $b=-0.7454$. Similarly, for oscillator 2, the scaling exponents for bounded chaos are $a=0.4578$ and $b=-0.7593$, and for EEs, they are $a=0.5153$ and $b=-0.7409$. From these power relations, we find that the scaling exponent ($a$) and power index ($b$) are approximately the same in the bounded chaos and EE cases. This shows that the trajectories of EE attractor grow from the bounded chaotic set.
To further elucidate our findings, we have plotted the power-law relationship (see Fig. \ref{powerICs}) for four different sets of initial conditions ICs: 
$IC_{1}=[x_{1}(0),y_{1}(0),x_{2}(0),y_{2}(0)]$ (yellow),
$IC_{2}=[-x_{1}(0),y_{1}(0),x_{2}(0),y_{2}(0)]$ (green), 
$IC_{3}=[-x_{1}(0),-y_{1}(0),x_{2}(0),y_{2}(0)]$ (orange),
and $IC_{4}=[x_{1}(0),-y_{1}(0),x_{2}(0),y_{2}(0)]$ (blue).
The scaling exponents of the power-law fit are computed as the mean of the exponents obtained from fits corresponding to different ICs, yielding , $a=0.5088$ and $b=0.7430$. The power-law fit (red line), along with the error variance (gray band), is shown in Fig. \ref{powerICs}. The fit approximates the data well in the extreme $|y_{max}|$ region, with some deviation observed at lower values of  $|y_{max}|$.
Finally, the established power-law relation from Figs. \ref{TSPWL}(c) and \ref{TSPWL}(f) and \ref{powerICs} shows a negative trend indicating that the magnitude of the velocity maxima increases as the conformal mass function $z$ decreases. For a given set of mass values, the corresponding $|y_{max}|$ can be approximately determined for different choice of ICs and appears to be a universal property of the Higgs oscillator.

\begin{figure*}[ht!]
    \centering
     \includegraphics[width=0.93\linewidth]{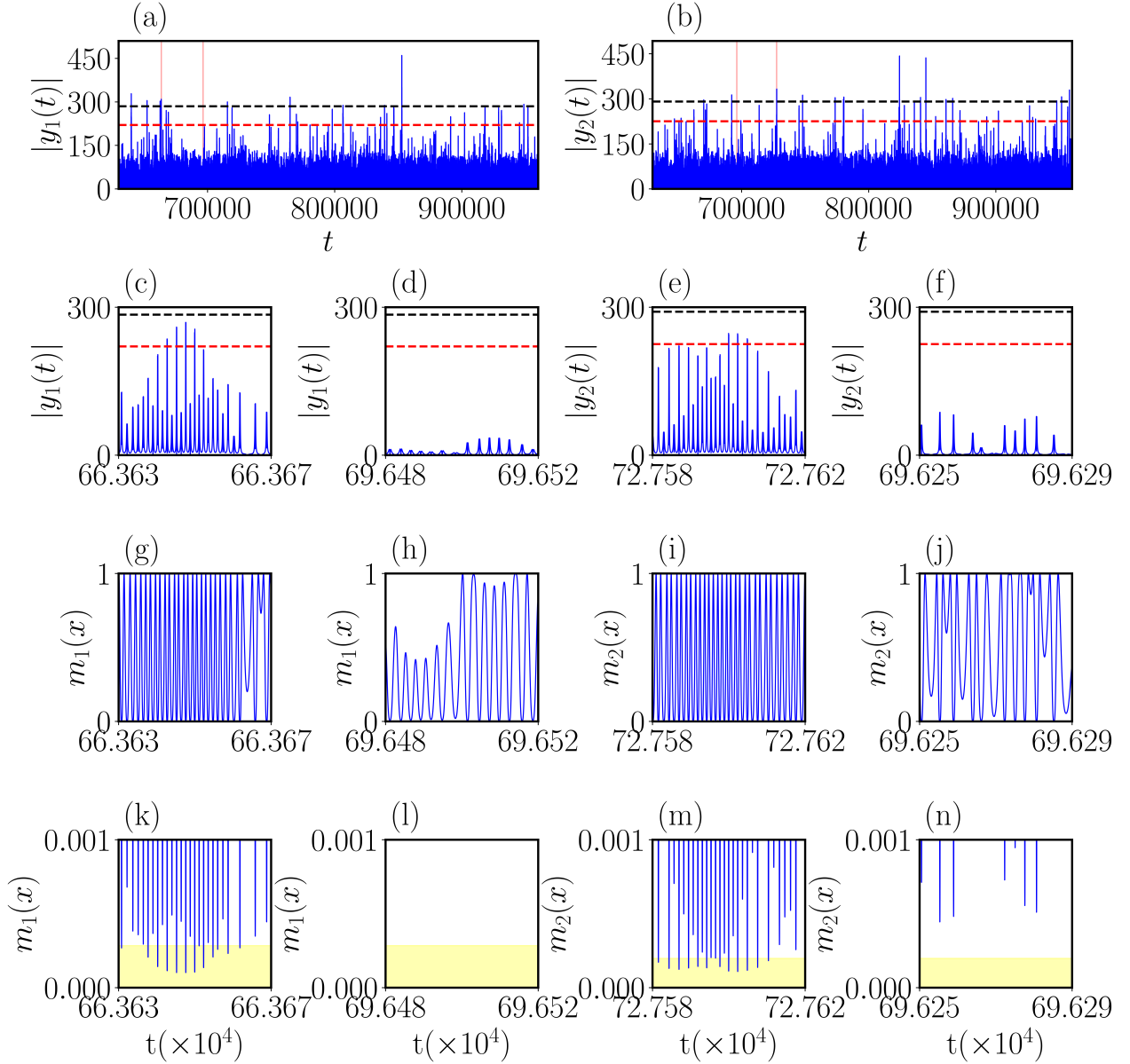}
     \caption{(a) and (b) show the time series of oscillators 1 and 2 for $(\epsilon = 13.1067)$. (c)–(d) and (e)–(f) present the zoomed views of the small vertically spanned red region for oscillators 1 and 2. The corresponding indicator variables $(m(x_{1}))$ and $(m(x_{2}))$ are shown in (g)–(h) and (i)–(j), with their zoomed versions in (k)–(l) and (m)–(n), respectively.
}
    \label{fig4}
\end{figure*}

The turning points (where acceleration $\ddot{y}(t)=0.0$) of the Higgs oscillator can be extracted and understood by the power law relations. Here, we have considered only the magnitude of the velocity extremes and it should be noted from the $m(x(t))$ vs $y(t)$ plot in Figs.~\ref{mass_velocity_phase_space}(a) and \ref{mass_velocity_phase_space}(c) that when the direction of velocity of the oscillator changes, the minima of $m(x(t))$ occur. Also, the variable $m(x)$ corresponding to the extreme event peaks are calculated using the power-law relation and they are represented as the green lines in the zoomed mass time series profile in Figs. \ref{mass_velocity_phase_space}(b) and \ref{mass_velocity_phase_space}(d). The corresponding $|y_{max}|$ occurs when the mass variation profile $m(x(t))$ intersects with critical mass (green line) which leads to the maximum of $y(t)$ (see Figs.~\ref{mass_velocity_phase_space}(b) and \ref{mass_velocity_phase_space}(d)). This confirms that whenever $m(x)$ reaches the critical conformal mass, the system exhibits extreme behavior. This can also be related to the early warning signal associated with the system (see Fig.~\ref{fig4}).

\subsection{Early warning indicator for N=2} 

So far, the mechanisms underlying extreme events (EEs) in coupled dynamical systems have primarily been investigated in terms of on--off intermittency and imperfect phase synchronization. In the present model, however, the oscillator possesses a position-dependent mass, and the established power-law relationship between $m(x)$ and $|y_{\max}|$ [Figs.~\ref{TSPWL}(c) and \ref{TSPWL}(f)] provides additional insight into the mechanism responsible for the emergence of EEs. This scaling suggests that complete synchronization can, in principle, be achieved when the conformal mass $m(x)$ is identical for all oscillators. Furthermore, owing to the parity symmetry of the conformal mass function,
$m(x)=\frac{1}{(1+\kappa x^2)^2}, ~\text{i.e.,}~ m(-x)=m(x),$
a uniform mass distribution may also lead to anti-phase synchronization. Therefore, to explain the genesis of extreme events in the coupled Higgs oscillator, we employ the synchronization error in the energy together with the established power-law relationship between the conformal mass and the maximum velocity.

From the two-parameter phase diagram shown in Fig.~\ref{two_phase}, we find that EEs predominantly emerge in the parameter regions characterized by asymmetric energy distribution among the oscillators, i.e., $\Delta H>0$, corresponding to energy-asynchronous states. Motivated by this observation and the formulated power-law relationship, we next adopt a suitable early warning indicator to signal when the bounded chaotic trajectory approaches the EE threshold.

So to find the critical value of the indicator variable $m^{*}(x)$ to signal the onset of EEs, we introduce the critical threshold value of $m^{*}(x)$  which can be calculated by the following two steps: 1) Take the bounded chaotic time series and calculate the EE like threshold $H_{m} = \langle y_{n}\rangle + 6.0*\sigma(y_{n})$, where $\langle y_{n}\rangle$ is the mean peak of velocity time series in the bounded chaotic set and $\sigma$ is the standard deviation of the set.  2) Since $H_{m}$ corresponds to the dimension of velocity $y$, we substitute $H_{m}$ in place of $y_{max}$ in the formulated power law relation. By doing a little algebra, we find the critical conformal mass $m^{*}(x)$ for the indicator threshold. The region of critical mass $m^{*}(x)$ is indicated in yellow color in the $m(x)$ time series plot in Fig.~\ref{fig4}.

To demonstrate the relationship, we first analyze the time series of the two oscillators (see Figs. \ref{fig4}(a) \& \ref{fig4}(b)). From these time series plots, we extract the velocity profile associated with extreme events (EEs) in four selected regions. Of these four regions, two exhibit bounded chaos (Figs. \ref{fig4}(d) \&  \ref{fig4}(f)), while the other two are characterized by extreme events (Figs. \ref{fig4}(c) \& \ref{fig4}(e)). Upon close comparison of the indicator variable time series with that of the velocity, it becomes evident that a maximum in velocity consistently appears whenever the indicator variable is reduced to its minimum (see Fig. \ref{fig4}(g - j)). Figures \ref{fig4}(k - n) provide zoomed-in versions of Figures \ref{fig4}(g - j) for more detailed observations. In Figures \ref{fig4}(k) and \ref{fig4}(m), the yellow shaded region corresponds to the mass-drop region, which is identified as the threshold for detecting extreme events. When the indicator variable falls on or above the yellow region, the velocity remains bounded under the extreme event threshold  (see and compare Figs. \ref{fig4}((d) \& (f)), \ref{fig4}((h) \& (j)) and  \ref{fig4}((l) \& (n))). Conversely, when the indicator variable enters the critical yellow region (the indicator threshold region), the trajectories cross or approach near the EE threshold (see and compare Figs. \ref{fig4}((c) \& (e)), \ref{fig4}((g) \& (i)) and  \ref{fig4}((k) \& (m))). This generally indicates that a smaller conformal mass, accompanied by increased disparity in the energy distribution between the oscillators, promotes higher particle velocities. Consequently, extreme events emerge when the conformal mass approaches very small values, whereas larger conformal masses do not support the occurrence of extreme events. 

\section{N-coupled Higgs oscillators} 

Next, we increase the number of the oscillators to $N=10$ and introduce a global coupling among the oscillators. One can also explore the collective dynamical states from the synchronization perspective by introducing a different coupling topology among the oscillators. Here, we focus ourselves in observing the EEs/SEEs and identifying the possible mechanisms in the globally coupled Higgs oscillator system.   
\begin{figure}
\centering
\includegraphics[width=0.75\linewidth]{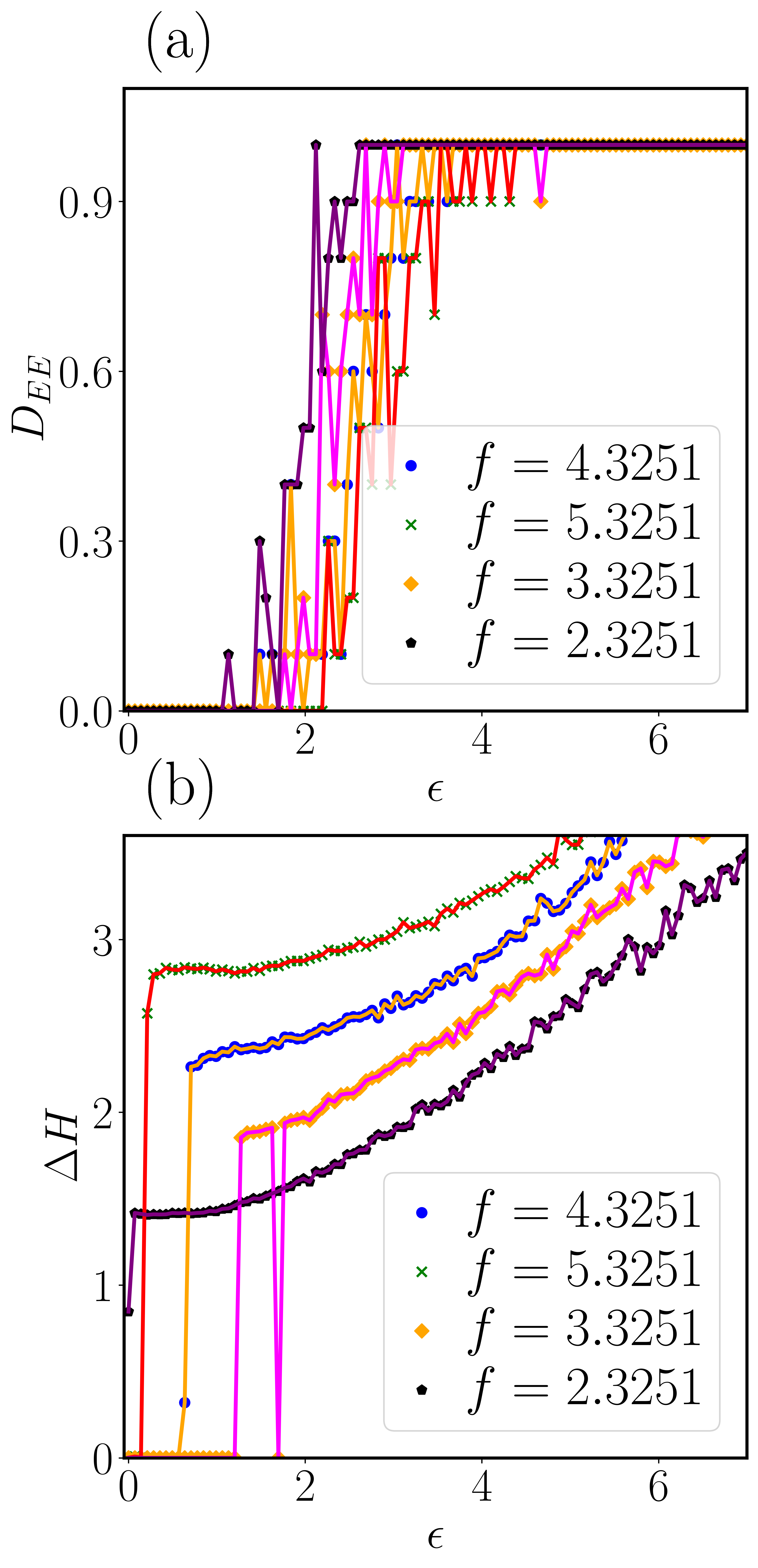}
\caption{Panel (a) represents the degree of oscillators $D_{EE}$ that transits to extreme events when scanning the interaction strength between the oscillators $\epsilon \in (0,7.0)$ for different values of f=[2.3251,3.3251.4.3251,5.3251] respectively. Similarly panel (b) represents the corresponding time averaged synchronization error of the instantaneous energy.}
\label{dee_10}
\end{figure}

\begin{figure}
\centering
\includegraphics[width=0.45\linewidth]{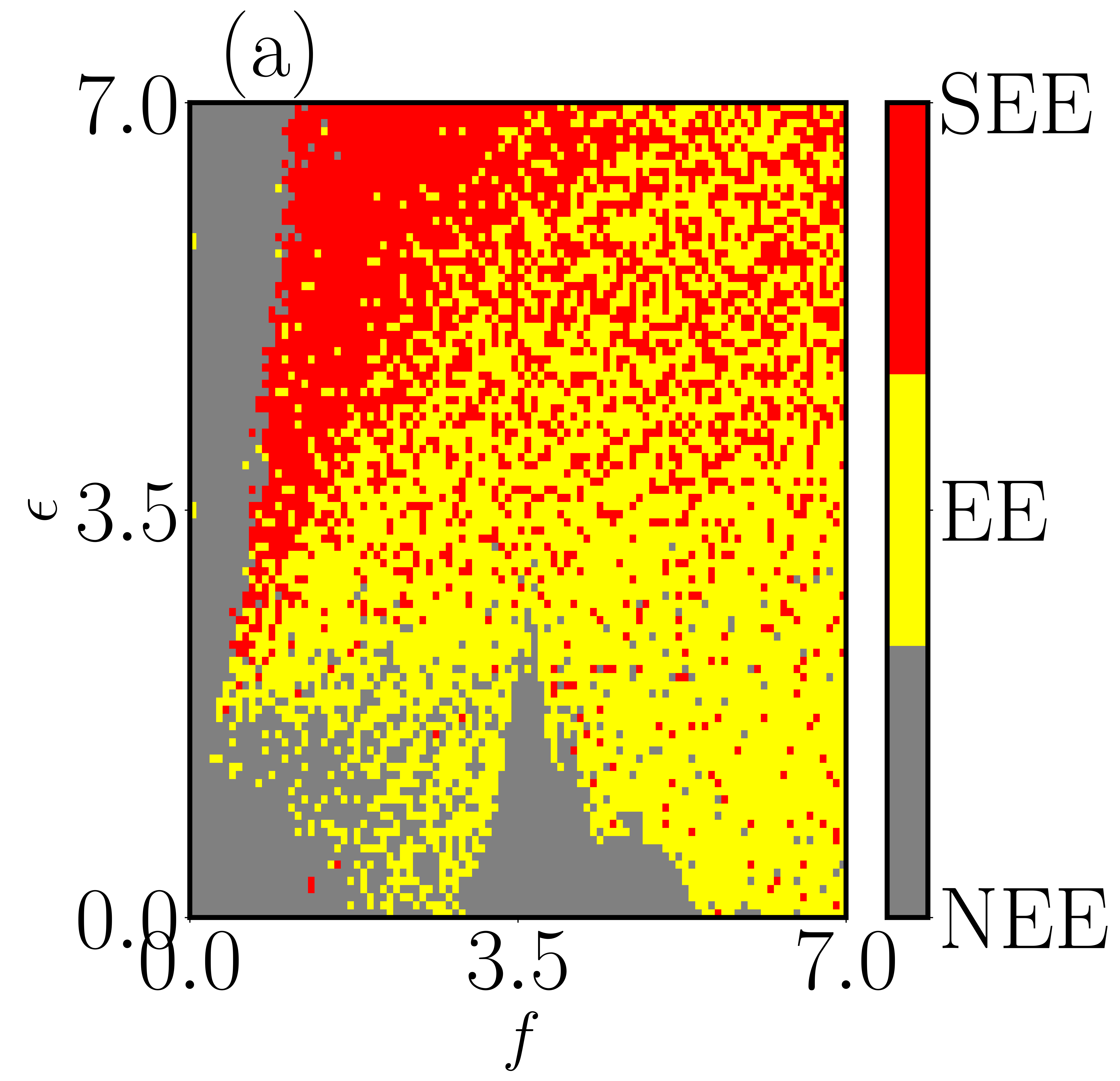}
\includegraphics[width=0.45\linewidth]{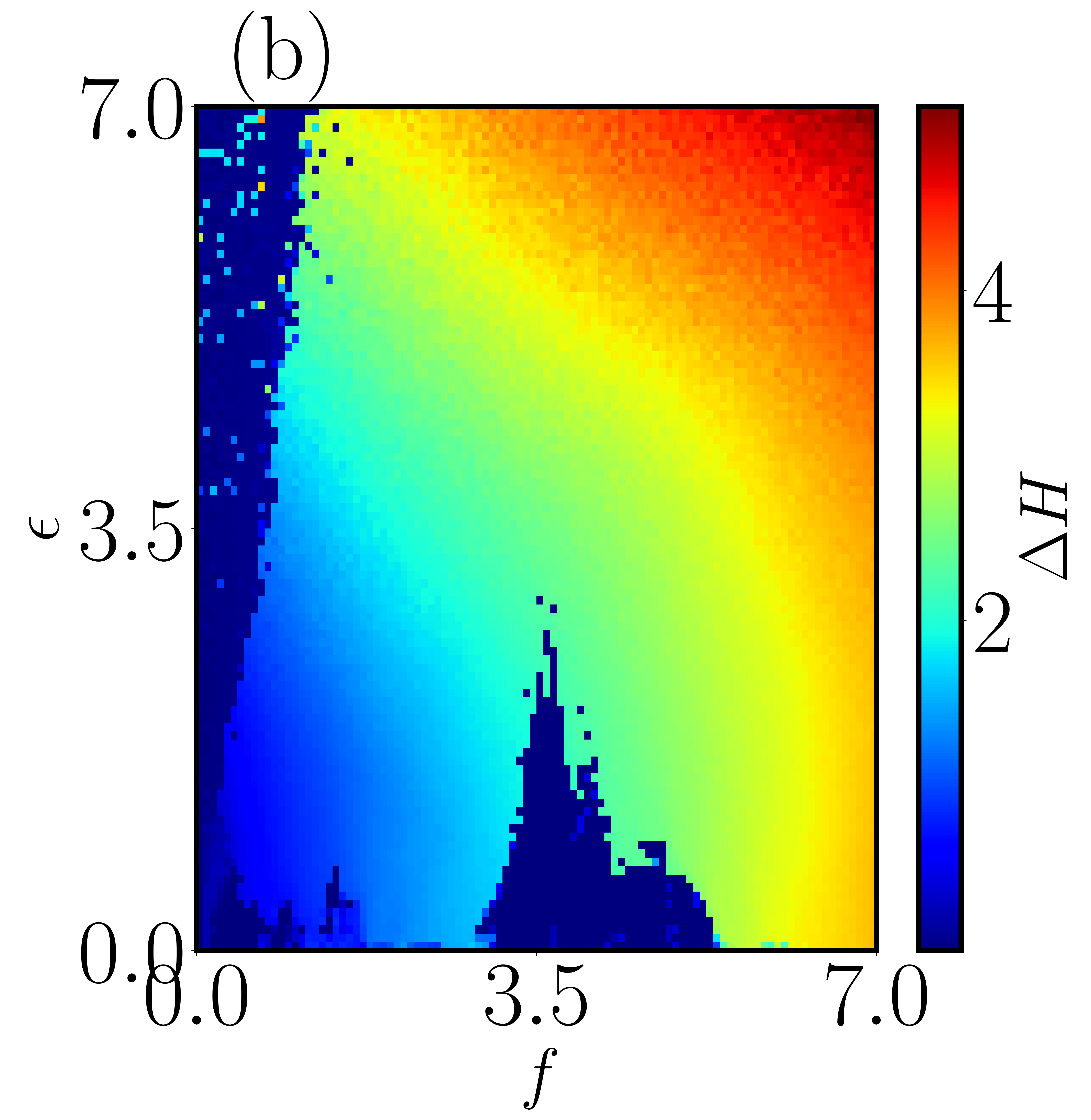}
\includegraphics[width=0.45\linewidth]{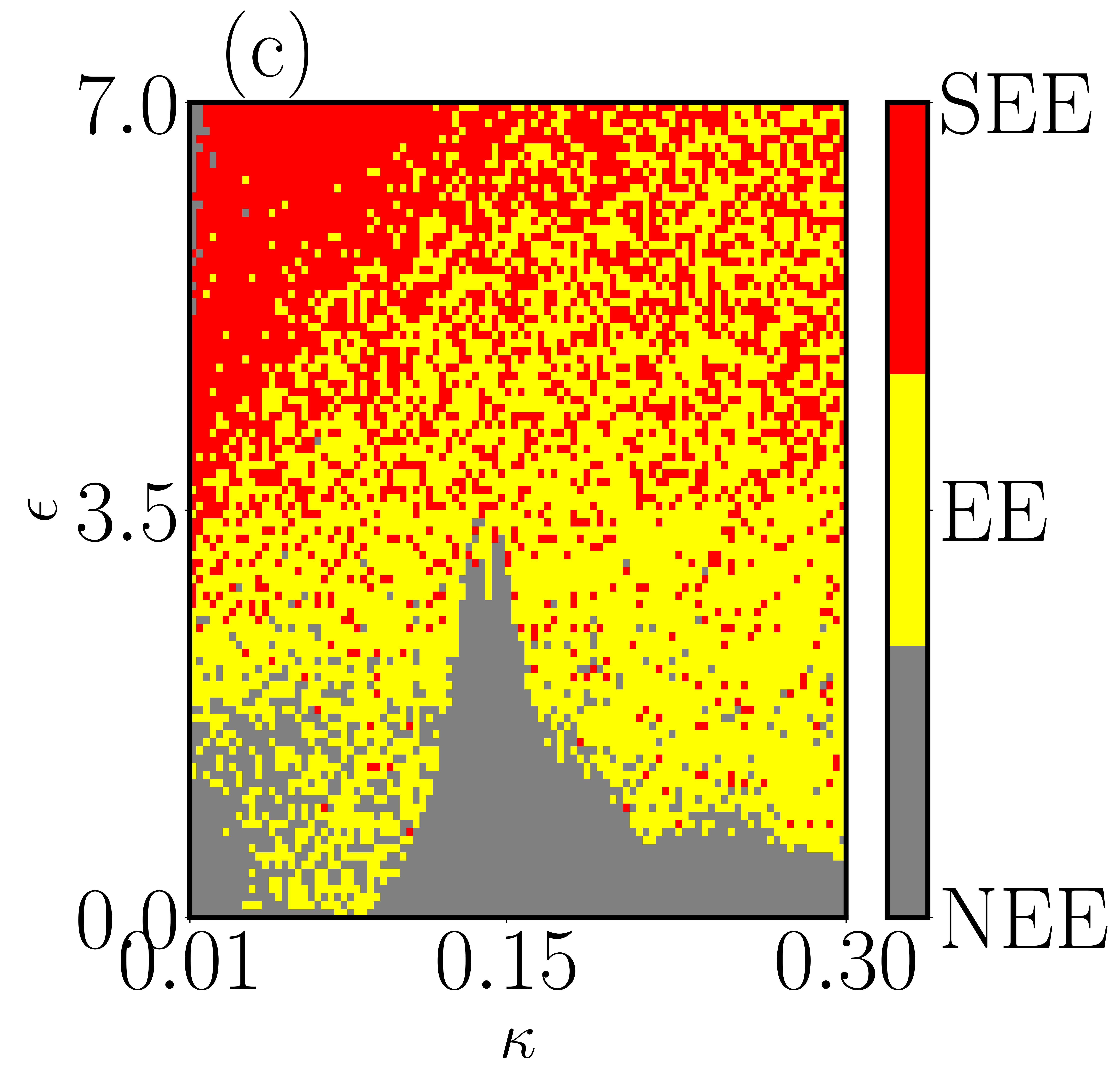}
\includegraphics[width=0.45\linewidth]{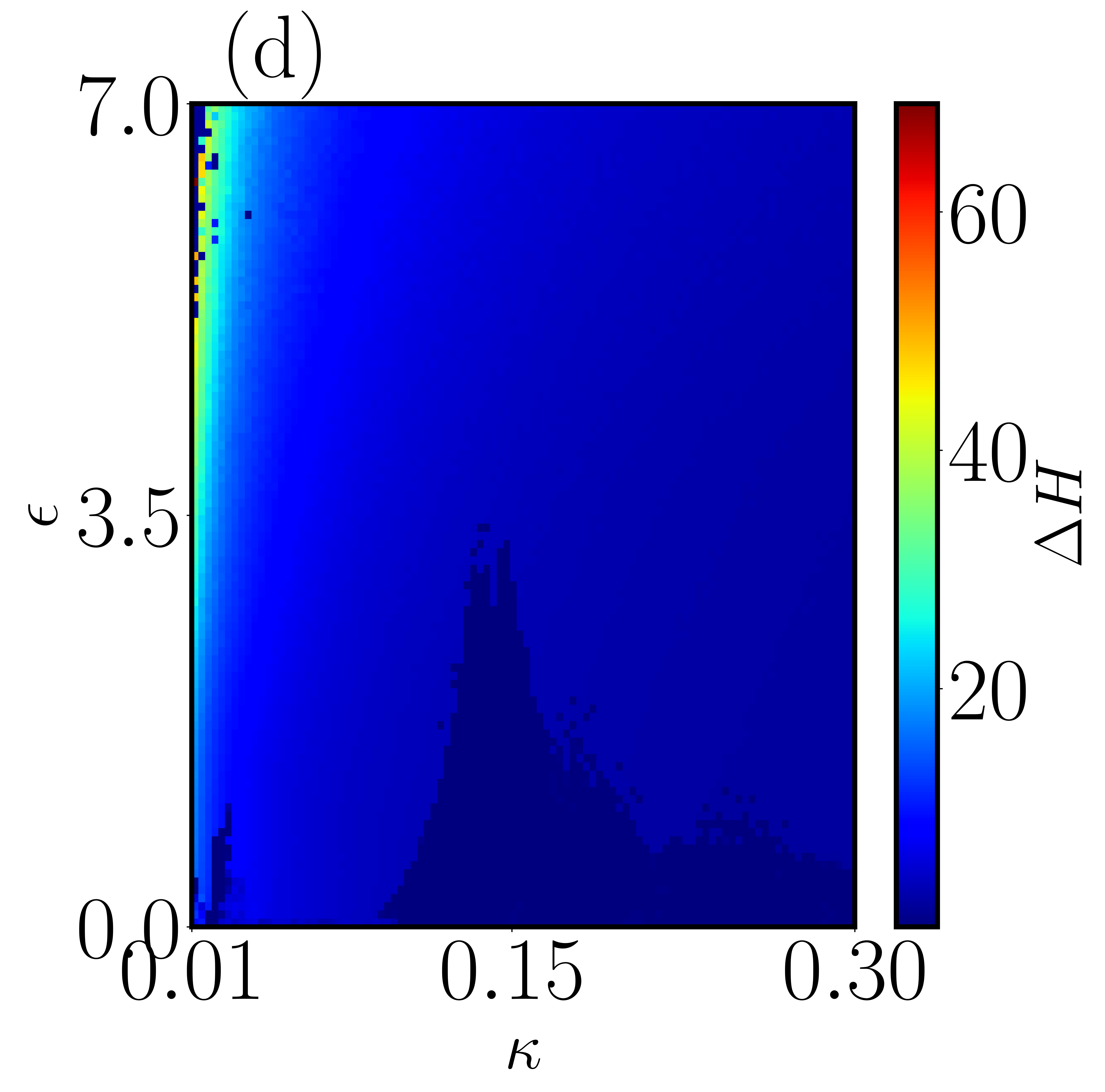}
\caption{(a) \& (c) Two parameter phase diagrams for ($\epsilon \in (0.0,7.0)$ vs $f \in(0.0,7.0) $) \& ($\epsilon\in(0.0,7.0)$ vs $\kappa \in(0.0,0.3) $) delineating the regions of non-extreme event (NEE)-grey, extreme event (EE)-yellow and super extreme event region(SEE)-red. Here (b)\& (d) are the corresponding time averaged synchronization errors between the instantaneous energy for N-coupled Higgs oscillator.}
\label{2phase_10}
\end{figure}

\begin{figure}
\centering
\includegraphics[width=0.6\linewidth]{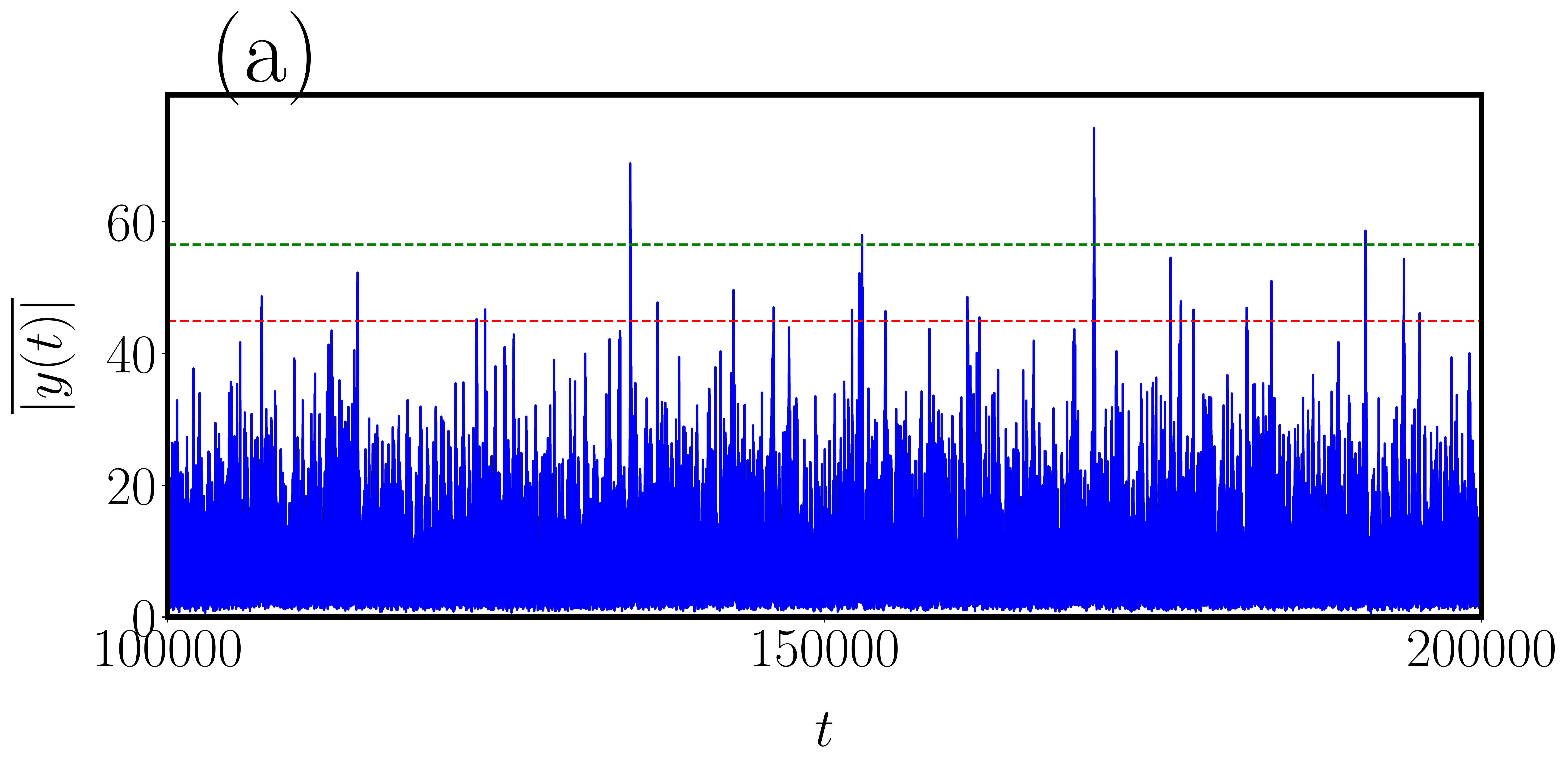}
\includegraphics[width=0.3\linewidth]{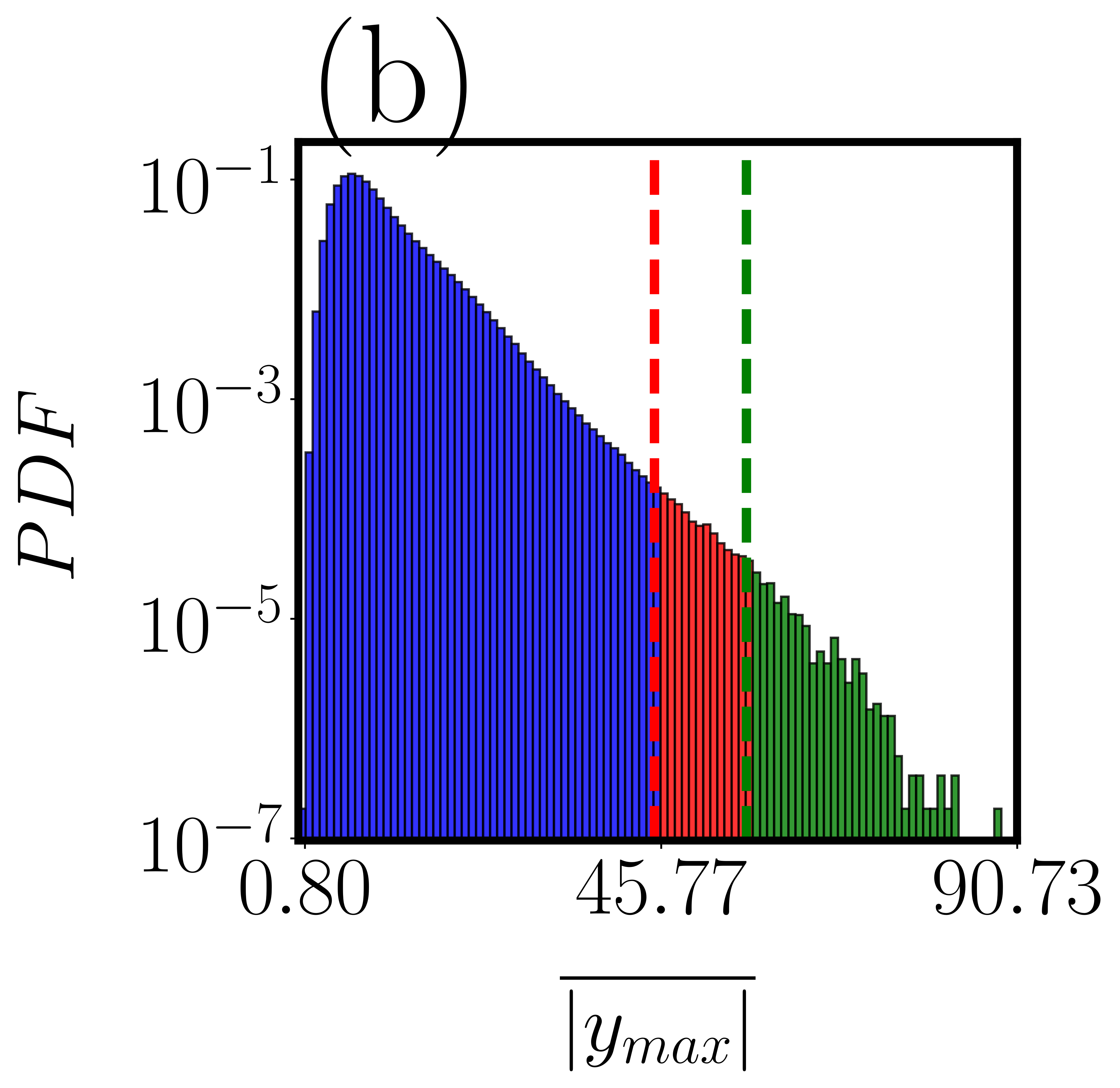}
\includegraphics[width=0.95\linewidth]{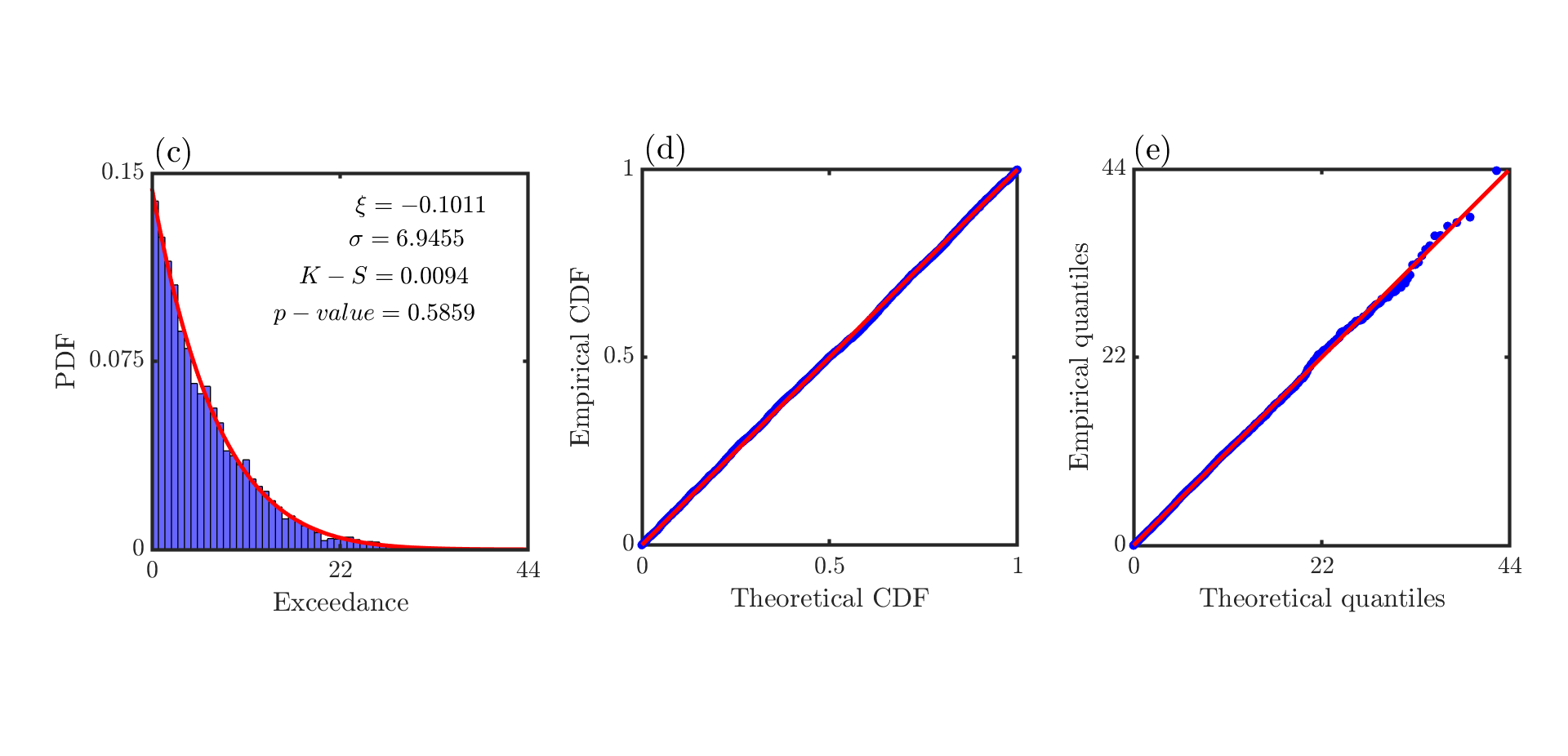}
\caption{(a) shows the extreme event (EE) associated  absolute collective average velocity profile time series for 10 interacting Higgs oscillators and (b) its PDF distribution for interaction strength value $\epsilon=4.7271$ for the fixed values of $\kappa=0.21$, $\omega_{o}^{2}=0.1$, $\alpha=0.00595$ , $f=1.3251$ and $\omega_{e}=0.54549$ respectively. Plots (c), (d) \& (f) represent the exceedance generalised Pareto distribution , P-P \&  Q-Q plots, respectively, which validate the tail behaviour and the EE threshold for characterising the EEs.}
\label{pdf_dee}
\end{figure}

\begin{figure}
\centering
\includegraphics[width=0.72\linewidth]{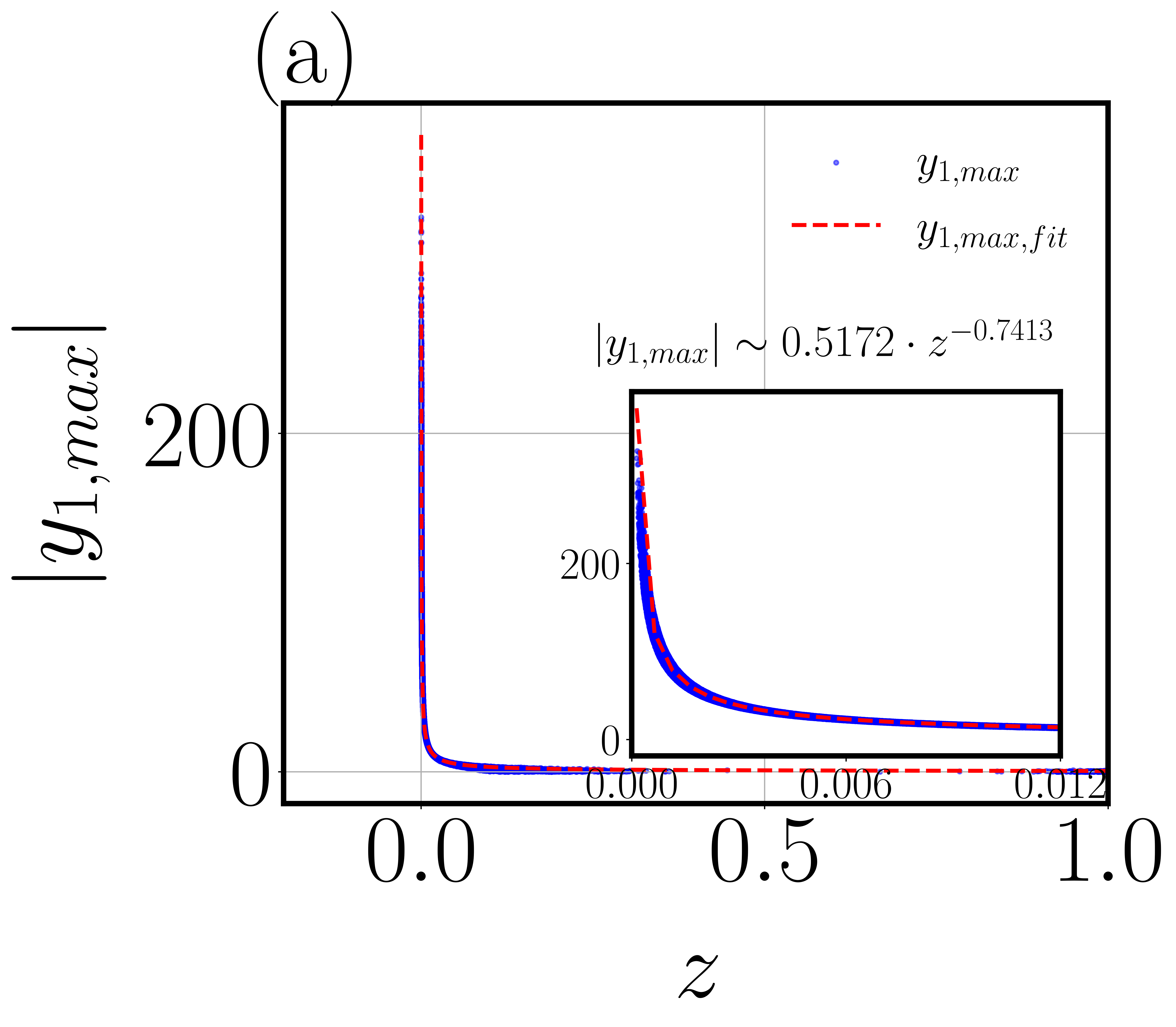}
\includegraphics[width=0.72\linewidth]{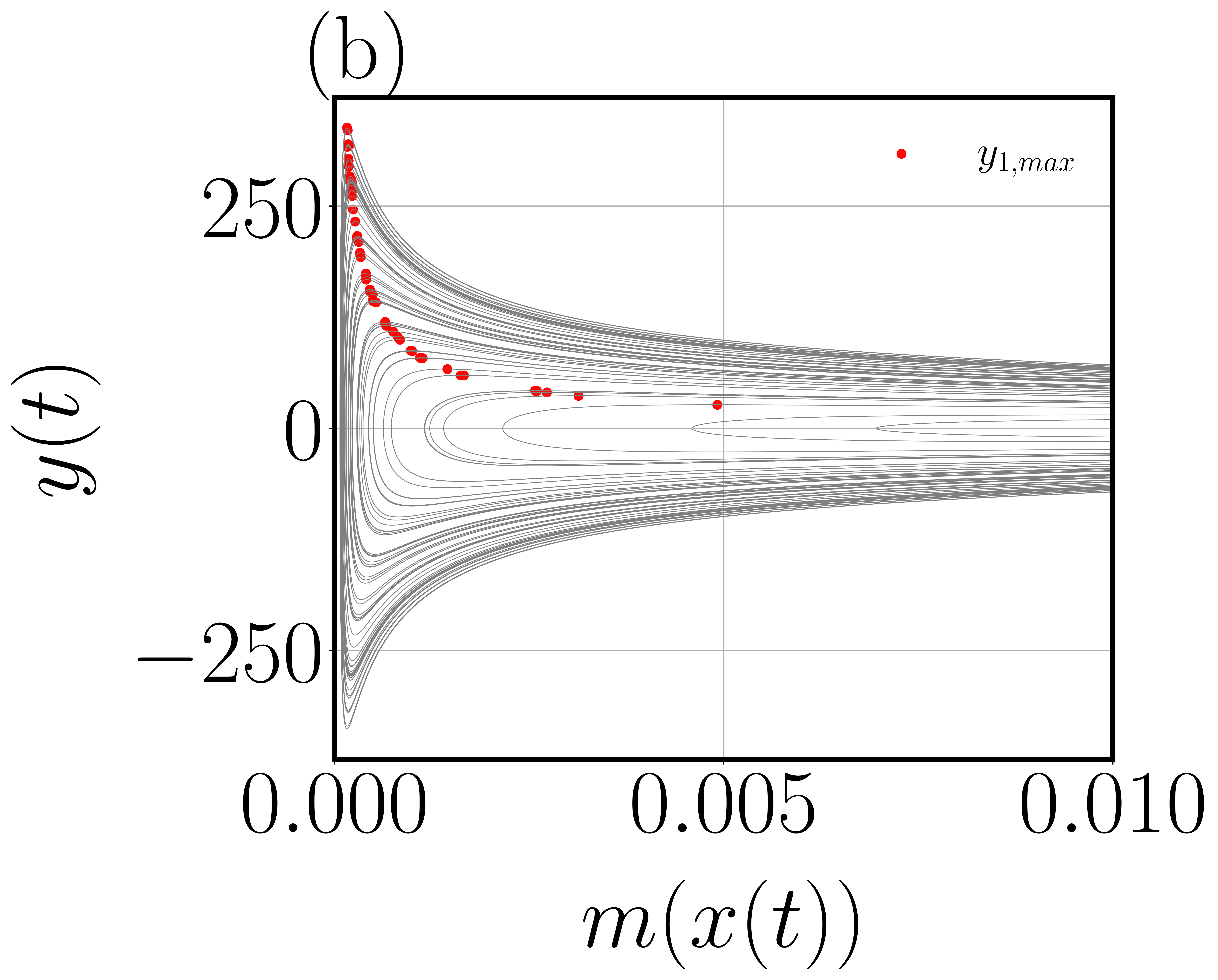}
\caption{(a) shows the power law relation of $|y_{1,max}|$ vs \textbf{$z$} for oscillator index 1 among the globally coupled $N=10$ oscillators for interaction strength value $\epsilon=3.5$. (b) shows the plot for $y(t)$ vs $m(x(t))$ and the replotted red dots represent the $y_{1,max}$. }
\label{power_law_Dee}
\end{figure}

For our analysis, we first calculate the degree of EEs  ($D_{EE}$) in the global network of the Higgs oscillators. The quantity $D_{EE}$ can be calculated by counting the number of nodes that exhibit EEs divided by the total number of nodes. In our analysis, we take $10$ globally coupled Higgs oscillators and study the dynamics by varying the interaction strength $\epsilon \in (0.0,7.0)$ for certain values of $f = [2.3251,3.3251,4.3251,5.3251]$ for the fixed set of parameters $\kappa=0.21$, $\omega_{o}^{2}=0.1$, $\alpha=0.00595$ and $\omega_{e}=0.54549$, respectively.

By varying the coupling strength $\epsilon$, we first analyze the fraction of nodes in the globally coupled system that undergo a transition to extreme events (EEs), while simultaneously monitoring the time-averaged energy synchronization error between the oscillators. For $f=2.3251$, EEs first emerge at $\epsilon=1.23$. As $\epsilon$ is increased further, the occurrence of EEs gradually spreads across the network, and at $\epsilon=2.31$ all the oscillators exhibit EEs, as shown in Fig.~\ref{dee_10}(a). Correspondingly, during the transition from non-extreme events (NEEs) to EEs, the energy undergoes an explosive transition from a synchronized to an energy-asynchronous state among the oscillators (see Fig.~\ref{dee_10}(b)). Notably, this explosive energy de-synchronization sets in before the critical coupling strength at which EEs first emerge (see Fig.~\ref{dee_10}). This observation further supports our conclusion that the emergence of extreme events in the coupled Higgs oscillator is preceded by an asymmetric distribution of energy among the oscillators. Repeating the analysis for progressively larger forcing amplitudes $f$, we find that the transition to EEs shifts towards higher values of $\epsilon$ (see Figs.~\ref{dee_10}(a) and \ref{dee_10}(b)). Moreover, beyond the explosive transition, the time-averaged Hamiltonian energy synchronization error increases steadily with increasing coupling strength $\epsilon$.

To elucidate the combined influence of the interaction strength $\epsilon$ and the forcing amplitude $f$ on the collective dynamics of the N-coupled Higgs oscillators, we construct the two-parameter phase diagram in the ($f,\epsilon$) plane by scanning $f \in(0.0,7.0)$ and $\epsilon \in (0.0,7.0)$. As shown in Fig.~\ref{2phase_10}(a), the system predominantly exhibits non-extreme events (NEEs - grey) for weak forcing amplitudes over the entire range of interaction strengths, with an additional narrow NEE region persisting around $\epsilon \approx 3.5$. In these regions, the corresponding time-averaged synchronization error $\Delta H$ remains nearly zero [Fig.~\ref{2phase_10}(b)], indicating an almost symmetric energy distribution among the oscillators. Outside these regions, the dynamical transition to extreme events (EEs - yellow) and super extreme events (SEEs - red), accompanied by a gradual radial increase in $\Delta H$ with increasing forcing amplitude $f$ and interaction strength $\epsilon$, can be observed. Furthermore, compared with the two-coupled oscillator system, the onset of extreme events in the $N$-coupled system occurs at relatively lower forcing amplitudes, highlighting that increasing the network size facilitates the higher energy interaction among the oscillators thereby robust emergence of collective extreme dynamics  (compare Fig.~\ref{two_phase}(a) with Fig.~\ref{2phase_10}(a)).

We further investigate the combined influence of the curvature parameter $\kappa$ and the interaction strength $\epsilon$ by constructing the two-parameter phase diagram in the ($\kappa,\epsilon$) plane, where $\epsilon \in (0.0,7.0)$ and $\kappa \in (0.01,0.30)$, as shown in Fig.~\ref{2phase_10}(c) and \ref{2phase_10}(d). We find that extreme events (EEs) and super extreme events (SEEs) persist throughout the region $\epsilon>2.15$ over the entire range of $\kappa$. In contrast, for $\epsilon <2.15 $, isolated regions of the parameter space exhibit non-extreme events (NEEs), where the corresponding energy synchronization error $\Delta H$ remains very small. Similar to the two-coupled Higgs oscillator, for very weak interaction strengths and across the entire range of $\kappa$, the oscillators exhibit an almost equal distribution of energy, as evidenced by the nearly zero values of $\Delta H$ [see Fig.~\ref{2phase_10}(d)].

A comparison between the two-coupled and N-coupled systems reveals that, for relatively large values of $\kappa$, the energy difference among the oscillators remains nearly uniform over the entire range of $\epsilon$, corresponding to the broad blue region in the phase diagram (see Fig.~\ref{2phase_10}(d)). In contrast, for smaller values of $\kappa$, the energy synchronization error increases progressively with increasing interaction strength, indicating a growing disparity in the energy distribution among the oscillators. These results demonstrate that the curvature parameter $\kappa$ plays a crucial role in maintaining a more uniform energy distribution as the coupling strength increases. Furthermore, increasing the number of coupled oscillators and the interplay of $\kappa$ results in a significantly larger energy synchronization error $\Delta H$ than that induced by variations in the forcing amplitude, as reflected by the corresponding color-bar ranges in Fig.~\ref{2phase_10}.

For the observation of EEs in the time series, we have fixed the interaction strength to be $\epsilon=4.7271$ and $f=1.3251$ with all other system parameters being fixed as mentioned above for $D_{ee}$ analysis of the N-coupled Higgs oscillator system. The mean average of the absolute velocity time series profiles of 10 globally coupled oscillators is presented in Fig.~\ref{pdf_dee}(a) and the corresponding PDF distribution is shown in Fig. \ref{pdf_dee}(b). To verify the tail behavior of the exceedance distribution for the mean absolute velocity of the oscillators, we fit a Generalized Pareto Distribution (GPD), obtaining the estimated shape and scale parameters $k = -0.1011$ and $\sigma=6.9455$, respectively. The negative value of the shape parameter $k<0$ indicates that the tail of the probability density distribution is bounded, implying the existence of a finite upper limit. The quality of the fit is supported by the P–P and Q–Q plots in Fig. \ref{pdf_dee}(d) \&  \ref{pdf_dee}(e), while the Kolmogorov–Smirnov test yields a statistic of 0.0094 with a $p-value$ of 0.5859, indicating excellent agreement between the empirical exceedances and the fitted GPD. These results validate the choice of the threshold parameter $\gamma=6$ for identifying extreme events.

To demonstrate the early warning scenario in the globally coupled oscillators ($N = 10$), we take any of the velocity time series among the $10$ oscillators that exhibit EEs. Here, we consider the EEs exhibiting oscillator index $1$ among the globally coupled ($N=10$) oscillators for demonstration. Similarly, from the preceding analysis of two coupled Higgs oscillators, we have plotted  $m(x_{1}(t))$ vs $y_{1}(t)$ (grey attractor) and replotted $y_{1,max}$ and their corresponding $m(x_{1,n})$ (red dots)  in Fig.~\ref{power_law_Dee} (b). On close observation in Fig.~\ref{power_law_Dee}(b), we note that the extremal velocity occurs when the value of $m(x)$ goes to zero. Similarly, the power-law relation and its exponents for oscillator 1 are given by $|y_{1,max}| \sim 0.5172 ~ z^{-0.7413}$.  From the previous observation in the two-coupled oscillator case, the limiting decay exponent for extremal velocity is approximately around $b \approx -0.8 \text{~to~} -0.75$ and the scaling exponent $a \approx 0.45 \text{~to~} 0.52$. Furthermore, it is to be noted that the increase in the scaling factor $a$ or decrease in the power exponent $b$ indicates the sudden and sharp expansion in the size of the attractor.

\section{Conclusion}

In this article, we investigated the nonlinear dynamics of two coupled Higgs oscillators interacting through a mass interaction term. While studying the dynamics of the oscillator, we encountered periodic, torus and chaotic solutions. On further tweaking the interaction strength among the oscillators, we found that the system exhibits extreme behavior in the velocity profile during the chronological time progression, identified using predefined EE threshold. The statistical properties of both bounded chaotic and extreme event dynamics are analyzed via PDF. The tail behavior of the extreme event distributions is further characterized by fitting a Generalized Pareto Distribution (GPD) curve to the distribution of exceedances. Further, to investigate the combined influence of the forcing amplitude $f$, the curvature parameter $\kappa$, and the interaction strength $\epsilon$ between the oscillators, two-parameter phase diagrams are constructed to delineate the regions of extreme events (EEs), super extreme events (SEEs), and non-extreme events (NEEs). The corresponding energy synchronization error, $\Delta H$, is also evaluated over the same parameter space to quantify the degree of energy sharing among the oscillators during the occurrence of EEs and NEEs. The genesis of extreme events in the coupled oscillators is elucidated by comparing the position-dependent mass profile with the velocity profile. To verify the underlying mechanism, we investigated the occurrence of EEs in a globally coupled Higgs oscillator network comprising $(N=10)$ oscillators. Our analysis demonstrates that the same dynamical mechanism identified for the two-coupled Higgs oscillator system remains valid for the larger network. Specifically, the genesis of EEs in the Higgs oscillator network~(\ref{eqn-coupled-n}) is governed by two essential factors: (i) an asymmetric redistribution of energy among the oscillators through coupling and (ii) a reduction of the position-dependent mass below a critical value, which facilitates large velocity excursions. The reduced critical mass acts as an early warning indicator for EEs. The conformal mass of the position-dependent mass system $m(x)$  arises purely from the geometry of curvature. Such a geometrical surface can be controlled by changing the potential surface or introducing additional perturbation when the oscillator reaches the critical conformal mass $m^{*}(x)$ to avoid extreme velocity $y_{EE}$ and to maintain the oscillator's velocity at an optimal level for system applications. We emphasize that the proposed power-law relationship is specific to the Higgs oscillator investigated in the present study and is established within the framework of this nonlinear system. Nevertheless, the methodology developed here can be readily extended to other nonlinear mechanical systems to examine whether analogous scaling laws exist. Exploring the existence of such scaling relationships across different classes of mechanical systems presents an interesting direction for future research.

\textit{Acknowledgments - } The author W.A.M acknowledges DST- INSPIRE, Govt. of India for the award of Junior Research Fellowship (DST INSPIRE – SRF IF210722/DST Inspire Fellowship/2022), A.V. acknowledges the DST-FIST for funding research projects via Grant No.SR/FST/College-2018-372 (C). M.L. wishes to thank the Department of Science and Technology for the award of a DST-ANRF National Science Chair under Grant No.  NSC/2020/000029 in which M.S is supported by a Research Associateship. V. C. would like to acknowledge SRM TRP Engineering College, India, for their financial support, Vide No. SRM/TRP/RI/005.


%

\appendix

\section{Additional Results}

\subsection{Two coupled oscillator : Two parameter phase diagram for $f$ vs $\epsilon$}

\begin{figure*}[h]
    \centering
   \includegraphics[width=0.7\linewidth]{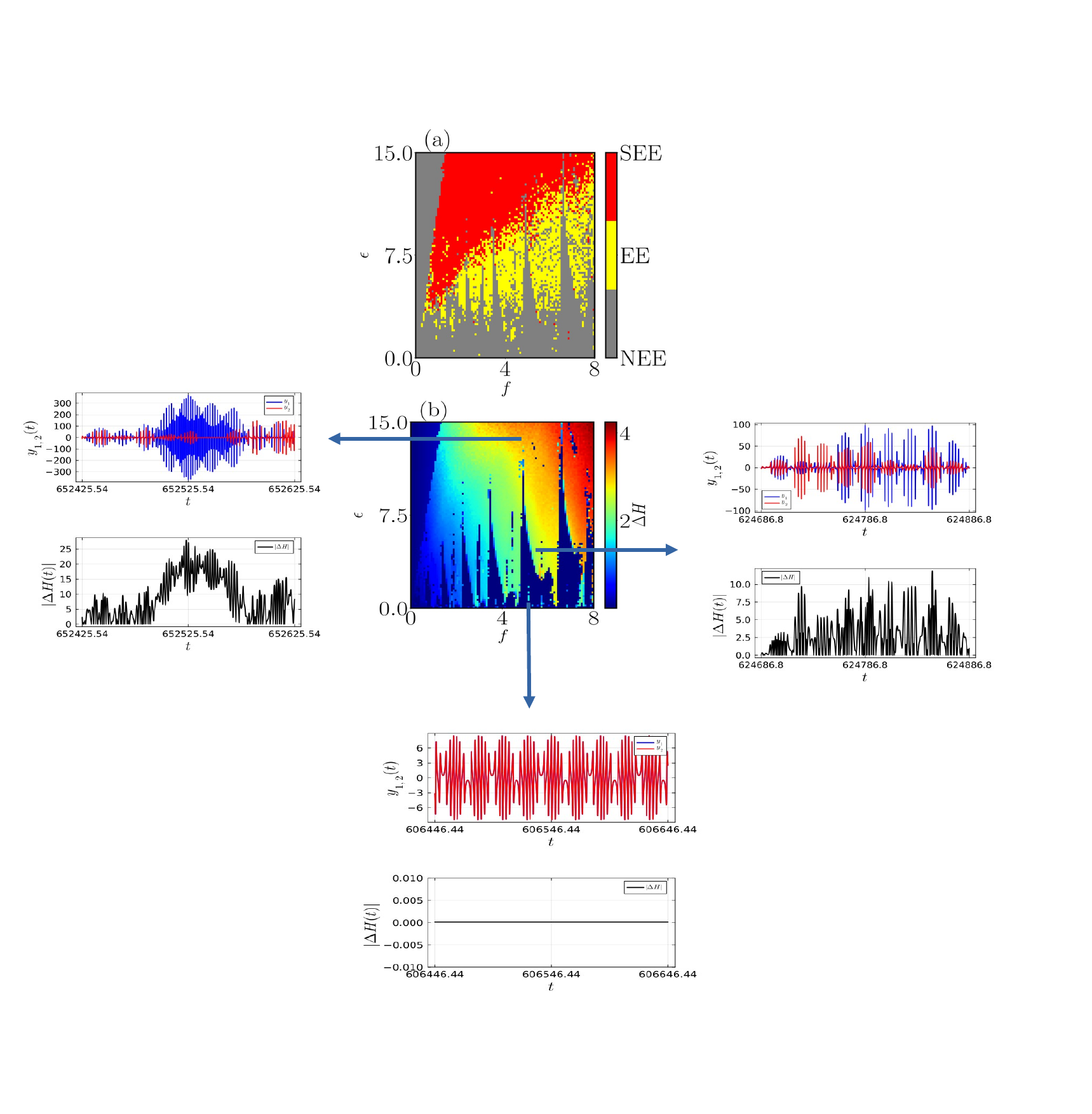}
 \caption{(a) Two parameter phase diagram for ($f\in (0.0,15.0)$ vs $\epsilon  \in(0.0,8.0) $) delineating the regions of non-extreme event (NEE) - grey, extreme event (EE) - yellow and super extreme event region(SEE)-red. Here (b) represents the corresponding time averaged synchronization error between the instantaneous energy for two-coupled Higgs oscillator system. Three representative regions (dark blue, light green, and red), indicated by arrows in panel (b), are selected to illustrate the replotted velocity profiles of oscillators 1 and 2, together with the corresponding time series of the instantaneous energy synchronization error.}
    \label{two_phase1}
\end{figure*}

In this Appendix, we consider three representative regions from the energy synchronization error plot (see Fig. \ref{two_phase1}). For each region, we compare the velocity time series of the two oscillators by replotting the trajectories of oscillator~1 (blue) and oscillator~2 (red). We also compute the corresponding instantaneous energy difference between the oscillators for the three selected regions indicated by the arrows in the energy synchronization error plot.

A comparison of the velocity time series with the energy difference shows that whenever one of the oscillators exhibits an extreme velocity, the energy difference between the two oscillators becomes significantly large. This behavior is clearly observed in the red region of the energy synchronization error plot, indicating that asymmetric energy pumping leads to a substantial transfer of energy to one oscillator. Consequently, the energized oscillator attains exceptionally large velocities.

In contrast, during the non-extreme events (NEEs), represented by the dark blue region in the energy synchronization error plot, the velocity profiles of the two oscillators remain synchronized, and the energy difference is nearly zero. This indicates symmetric energy pumping, where the energy is distributed almost equally between the oscillators.

For the bounded chaotic regime, the oscillators exhibit a finite energy difference, indicating the presence of asymmetric energy pumping. However, its intensity is considerably weaker than that observed in the super-extreme-event regime, resulting in bounded chaotic oscillations rather than exceptionally large velocity excursions.

The same energy error dynamics is applicable to $n$-coupled oscillators and to the interplay of different control parameters.

\end{document}